\documentclass[12pt]{article}   
\usepackage{epsfig}

\newcommand{\locsection}[1]{\setcounter{equation}{0}\section{#1}}
\renewcommand{\theequation}{\thesection.\arabic{equation}} 

\def\beq{\begin{equation}}
\def\eeq{\end{equation}}
\def\pl{\partial}

\def\al{\alpha}

\def\ga{\gamma}

\def\de{\delta}
\def\De{\Delta}

\def\Si{\Sigma}
\def\te{\theta}
\def\tm{\times}
\def\La{\Lambda}
\def\lam{\lambda}

\def\om{\omega}

\def\sq{\sqrt}

\def\l{\left (}
\def\r{\right )}
\def\fr{\frac}
\def\la{\label}
\def\hs{\hspace}

\def\ran{\rangle}
\def\lan{\langle}
\def\ov{\overline}
\def\tl{\tilde}
\def\hsp{\hspace}

\begin{document}
\begin{titlepage}

\begin{center}   
{\Large\bf 
Gauge Coupling Unification and Phenomenology \\   
\vspace{0.2cm} 
of Selected Orbifold 5D $N=1$ SUSY Models}  
\end{center} 
\vspace{0.5cm} 
\begin{center} 
{\large Filipe \hs{-0.07cm}Paccetti \hs{-0.07cm}Correia$^{a}$, 
{}Michael \hs{-0.07cm}G. \hs{-0.07cm}Schmidt$^{a}$, 
{}Zurab \hs{-0.07cm}Tavartkiladze$^{a, b}$
} 
\vspace{0.5cm}

$^a${\em Institut f\"ur Theoretische Physik, Universit\"at Heidelberg,
Philosophenweg 16,\\
 D-69120 Heidelberg, Germany \\

$^b$ Institute of Physics, 
Georgian Academy of Sciences, Tbilisi 380077, Georgia\\

} 

\end{center}

\vspace{1.0cm}

\begin{abstract}

We study gauge coupling unification and
various
phenomenological issues, such as baryon number conservation, the $\mu $
problem and neutrino anomalies, within SUSY 5D orbifold models. The 5D
MSSM on an $S^{(1)}/Z_2$ orbifold with 'minimal' field
content does not lead to low scale unification, while some of its
extensions
can give unification near the multi TeV scale. Within the orbifold $SU(5)$
GUT,  low scale unification can not be realized due to full
$SU(5)$ multiplets participating in the renormalization above the
compactification scale. As alternative examples, we construct 5D $N=1$
SUSY Pati-Salam $SU(4)_c\tm SU(2)_L\tm SU(2)_R\equiv G_{422}$ and flipped 
$SU(5)\tm U(1)\equiv G_{51}$ GUTs [both maximal subgroups of $SO(10)$]
on an $S^{(1)}/Z_2\tm Z_2'$
orbifold.  New examples of
low scale unifications within $G_{422}$ are
presented. For $G_{51}$ the  unification scale is shown 
to be necessarily close to $\sim 10^{16}$~GeV.
The possible
influence of brane couplings on the gauge coupling unification is
also outlined. 
For the resolution of the various phenomenological problems 
extensions
with a discrete ${\cal Z}$ symmetry turn out to be very effective.

\end{abstract}
\end{titlepage}

\section{Introduction: Old and new features of GUTs
}

The standard model of elementary particle physics (SM) gives an
excellent explanation of all existing experimental data. However, there
are
quite strong theoretical motivations to believe that the SM
is an effective theory of a more fundamental theory and that the gauge
couplings of
$SU(3)_c\times SU(2)_L\times U(1)_Y\equiv G_{321}$ have a common origin.
The construction of grand unified theories
(GUTs) \cite{guts}, which unify $G_{321}$ gauge interactions in a single
non
Abelian group [$SU(5)$, $SO(10)$, $E_6$ etc], give an elegant explanation
of charge quantization and also unify quark-lepton families.
The idea of GUT got a great support from the fact that the
three gauge couplings measured at that early times were indeed unifying
at
energies near $M_G\sim 10^{15}$~GeV \cite{nsuniGUT}. 
Progress in measuring the strong gauge coupling and also the weak
mixing angle $\sin^2 \te_W$ with higher accuracy has ruled out the
minimal $SU(5)$ GUT [and also minimal $SO(10)$ without intermediate
scale] from the viewpoint of coupling unification
\cite{nonunifSM}, \cite{nonunifMSSM}. However, the
minimal
supersymmetric extension of the standard model (MSSM) and also the minimal
SUSY $SU(5)$ GUT \cite{susyGUTs}(which except for GUT threshold
corrections both have the 
same pattern 
of running of couplings below $M_G$)
were giving  values for the $\al_3(M_Z)$ coupling 
\cite{unifSUSY1}-\cite{nonunifMSSM} well within the
experimental limits at that time. 
Indeed since SUSY theories stabilize
hierarchies, for realistic model building supersymmetry might be
the best way to proceed.
It is assumed, that the SUSY breaking scale $m$ lies in a range 
$500$~GeV - few TeV
and below this characteristic scale the theory is just the SM with minimal
particle content except the Higgs sector, while above the $m$ scale the
theory is
supersymmetric.
Despite these nice features of SUSY theories, there are
various
puzzles and problems, which are connected with SUSY GUTs and we will list
some of them here.

{\bf (i)} Baryon number violation is a particular feature of GUTs such as
$SU(5)$, $SO(10)$. Since for SUSY GUTs the unification scale
$M_G\simeq 2\cdot 10^{16}$~GeV is larger than for non SUSY GUTs, the gauge
mediated $d=6$ nucleon
decay is compatible with the latest SuperKamiokande (SK) limit
$\tau_N \stackrel{>}{_\sim }10^{33}$ yrs \cite{pdg}. However, with SUSY
there is a new source for nucleon decay through $d=5$ operators, which
makes the minimal $SU(5)$ and $SO(10)$ scenarios incompatible 
\cite{pdecaySU5} with SK data.

{\bf (ii)} The unified multiplets of minimal $SU(5)$ lead to the wrong
asymptotic mass relations $\hat{m}_d^{(0)}=\hat{m}_e^{(0)}$.
In the minimal $SO(10)$ the situation is even worse, since  
$\hat{m}_u^{(0)}=\hat{m}_d^{(0)}=\hat{m}_e^{(0)}$ and 
$\hat{V}_{CKM}={\bf 1}$ is predicted.

{\bf (iii)} 
The problem of doublet-triplet (DT) splitting in the Higgs supermultiplet 
still needs to
be resolved. In GUTs, the MSSM Higgs doublets are usually accompanied by
colored triplets. In order to maintain coupling unification and reasonably
stable nucleons [triplets could induce nucleon decay through $d=5$
operators, see {\bf (i)}], triplet components must be superheavy. So, one
should provide a natural explanation of the fact that sometimes states
(Higgs
doublets and colored triplets) coming from the same GUT multiplet are
split with a huge mass gap $M_T/M_D\geq 10^{13}$.

{\bf (iv)} 
The spontaneous breaking of the GUT symmetry requires scalars in a high
representation of the  gauge group considered; thus the superpotential,
responsible
for
symmetry breaking, contains many unknown parameters and usually looks
rather
complicated.

{\bf (v)} 
The so-called $\mu $ problem exists even within the MSSM. 
4D superpotential couplings allow a $Mh_uh_d$ term,
where $h_u$, $h_d$ are the MSSM Higgs doublets and $M$ is some mass close
to the cutoff scale of the theory. So, somehow large values for $\mu
$($\sim
M$) must be avoided.  
In order to have the desired electroweak symmetry breaking and
a reasonable phenomenology, a $\mu $ term of the
magnitude $\sim 500$~GeV - few TeV has to be generated in a good model
(also within GUTs after solution of the DT splitting problem 
{\bf (iii}) and having succeeded to obtain $\mu =0$ ).

{\bf (vi)} Recent atmospheric \cite{atm} and solar \cite{sol} neutrino SK
data have confirmed neutrino oscillations. The explanation of the
atmospheric anomaly (by a characteristic mass squared scale 
$m^2_{atm}\sim 10^{-3}$~eV$^2$) already forces us to step beyond the
MSSM and the minimal SUSY $SU(5)$ (a neutrino mass  $\sim 10^{-5}$~eV can
be generated through Planck scale $d=5$ operators and can explain the
solar anomaly through large angle vacuum oscillations. However, this
solution is disfavored by the SK data). In order to have neutrinos with
masses 
$0.1-1$~eV, the lepton number must be violated by a proper amount.
This requires considering extensions of the MSSM and the minimal SUSY
$SU(5)$. 
It would be most
welcome if the considered model would contain a source  
for the needed lepton number violation.  

{\bf (vii)} Very accurate measurements of $\al_3(M_Z)$ \cite{pdg} already 
allow to judge whether a given GUT scenario is viable or not. Two loop
renormalization studies of the MSSM (with all SUSY particles near the
$M_Z$
scale) predict $\al_3(M_Z)=0.126$ \cite{nonunifMSSM1},
which contradicts the
experimental $\al_3^{exp}(M_Z)=0.119\pm 0.002$ \cite{pdg}. This situation
can be
improved either by pushing all SUSY particle masses up to the $\sim 3$~TeV
mass scale \cite{nonunifMSSM1}, or by some GUT threshold corrections.
With the latter the minimal SUSY
$SU(5)$ does not give any promising results \cite{nonunifSUSYsu5}.
Comparing GUT
scenarios, those would be considered more attractive 
which, without constraining the
SUSY particle mass spectra, give acceptable values for the strong
coupling.

On the theoretical side quite a few new possibilities have been
found since the early days of GUTs and also the introduction of SUSY.

{\bf $\al $)}
String theory is primarily a theory of (super)gravity
but it also contains in a less unique manner matter and gauge fields.
It had an enormous effect on the taste of model builders although concrete  
phenomenological results are still not obvious. We particularly mention
symmetry breaking mechanisms not requiring very high Higgs representations
of the GUT gauge groups and a natural assignment of fundamental
representation
to matter. Also the possibility to calculate (in principle!)
Yukawa couplings is very impressive. But unfortunately there is a huge
and even increasing number of string vacua with (presently) no possibility
to make a choice of one or the other except for phenomenological
reasons. Still until recently \cite{branint} it was notoriously difficult
to find
a string model realization implementing the SM with three generations.

{\bf $\beta $)}
Extra dimensions: The Kaluza-Klein use of extra
dimensions to be curled up one way or the other had several
renaissances. Of course it is very tempting to obtain extra model
informations from some extra dimensions - there is plenty of space
in this dreamland, which can contain geometry/topology. In string
theory extra dimensions are mandatory for consistency. A drawback then seems
to be that such extra dimensions would show up only at the string scale
which is normally identified with the Planck scale of our gravity. One then
is led to talk about physics which presumably never will be tested
in the laboratory. Recently it became a point of common interest whether 
the
string scale might be as low as the TeV scale \cite{lowstr} still
allowing for our
gravity scale. In this case higher dimensions should show up soon in
experiments \cite{lowstrpen}.

{\bf $\gamma $)}
It was exciting news that dualities connect the various
types of string theories \cite{strings}. 
The open string picture allows
for $D$-branes which contain our 3-dimensional space but also allow
for some extra dimensions which may be curled up or projected out in the
case of intersecting branes. This version of string theory may be
particularly appropriate also in the case of singular points of divided 
out symmetries for an approximation by a description in local
quantum field theory language since there are no winding states.

In resolving  problems {\bf (i)-(iv)} of SUSY GUT scenarios the orbifold
constructions seem to be very promising \cite{kawa}-\cite{orbSO10}. In
the original
paper of ref. \cite{kawa}, a five dimensional
(5D) $N=1$
SUSY $SU(5)$ GUT on an $S^{(1)}/Z_2\times Z_2'$ orbifold was considered. 
Due to this
construction, it turns out that the problems {\bf (i)}-{\bf (iv)} can
be
resolved in a very natural way for a wide class of unified models 
\cite{kawa}-\cite{orbSO10}, while {\bf (v)}-{\bf (vii)} still depend on
peculiarities of the scenario considered and will be discussed 
in more detail below. Due to specific boundary conditions, 
it is
possible to mod out selected sub-states from a given GUT representation. 
Through this self consistent procedure, it is possible to obtain the
desired GUT symmetry breaking, nucleon stability 
and natural DT splitting.

In the last years, theories with extra dimensions have attracted great
attention. Originally the main phenomenological motivation was the
possibility to
resolve the gauge hierarchy problem without supersymmetry. It was observed
\cite{tevgr1}, that due to sufficiently large extra dimensions, it is
possible
to lower the fundamental scale $M_f$ even down to a few TeV (indeed, this
can
be an excellent starting point for understanding the
electroweak scale), while the 4D Planck mass still has the required value 
$\sim 10^{19}$~GeV. Due to the large extra dimensions,  Newton's law
could be modified at short distances
where the behavior of gravity is
still unknown and is studied in ongoing experiments \cite{gravexp}. 
Similarly and perhaps with a richer phenomenology 
\cite{lowstrpen},
one can study the spectrum for scenarios with a string scale of a few TeV
\cite{lowstr}. 
It turned
out, that the presence of extra dimensions can play a crucial role also
for
obtaining low scale unification of gauge couplings 
\cite{lowunif1}-\cite{lowunif4} through
power law running \cite{ven}. The construction of realistic GUT scenarios
with 
low scale unification raises the  hope that 
phenomenological implications can be detected. However, the
orbifold GUT scenarios considered up to now
do not allow for low scale unification \cite{orbunif},
\cite{orbunif1} because in these
settings
the GUT symmetry is restored at energies higher
than the compactification scale. Thus 
full GUT
multiplets [either of $SU(5)$ or $SO(10)$]
will participate in the running
and power law unification does not take place. 
Relatively low scales $\sim
10^{13-14}$~GeV are also preferable for lepton number violation. One way
for obtaining unification on a scale much below $\sim 10^{16}$~GeV is
to consider either GUT models with product groups  or with
(intermediate) stages of symmetry breaking - step by step compactification 
of more than one extra dimension. On
the GUT scale $M_G$ a first
step
compactification ($1/R'\sim M_G$) takes place and the unified group $G$
reduces
to its
subgroup $H$. In the second step compactification, whose scale  
$\mu_0=1/R$ is below $M_G$,  the subgroup $H$ is broken.
If $H$ is different from $SU(5)$ and if the states are 
non complete multiplets of $SU(5)$, then due to their contribution to the
running
between $\mu_0$ and $M_G$ there can appear 
power law unification on intermediate
or low scales.

In this paper we consider 5D $N=1$ SUSY models with orbifold
compactifications. We start our discussion with the standard model
$G_{321}$ gauge group and an $S^{(1)}/Z_2$ orbifold.
In order to have a model without the 
phenomenological
problems {\bf (i)}-{\bf (vi)} [in this case except the {\bf (ii)}-{\bf
(iv)} of course], we
introduce a discrete ${\cal Z}$ symmetry which elegantly resolves problems
{\bf (i)}, {\bf (v)} and  replacing  matter $R$ parity
allows for some lepton number violating couplings which can generate
neutrino
masses. Thus, {\bf (vi)} also can be resolved. 
We confirm that, with
the MSSM states plus appropriate Kaluza-Klein (KK) excitations,
successful unification holds only for $1/R\simeq M_G\sim 10^{16}$~GeV. 
Low scale unification requires either some extensions 
\cite{lowunif2}, \cite{lowunif3} 
or the existence of specific threshold corrections \cite{lowunif4}. 
Similarly we can discuss 
5D $N=1$ SUSY $SU(5)$ GUT on an $S^{(1)}/Z_2\times Z_2'$ orbifold. 
In this setting the problems {\bf (ii)}-{\bf (iv)} are resolved naturally,
while {\bf (i)}, {\bf (v)} will again be resolved by introducing a
discrete
symmetry ${\cal Z}$.
As far as the gauge coupling unification is concerned,  all states
including  matter
supermultiplets and their copies
form full $SU(5)$ multiplets above the compactification scale. Because of
this, low and intermediate
scale unification can not take place. 
Then we address the question whether  power law unification is possible or
not 
(at low or intermediate scale) within the orbifold GUT construction. 
We emphasize the possibility of a so-called step by step compactification
with an
intermediate gauge group, different 
from the $SU(5)$ in structure and field content. 
This potentially allows for power law
unification. Besides the latter a quite  different and peculiar
phenomenology
can arise.
To demonstrate this we consider Pati-Salam
$SU(4)_c\times SU(2)_L\times SU(2)_R\equiv G_{422}$ \cite{PS} and flipped 
$SU(5)\times U(1)\equiv G_{51}$ GUTs. Both these gauge groups are maximal
subgroups 
of $SO(10)$  
\cite{SU51ofSO10}, \cite{slansky} and thus one could imagine that they
are produced in a first step  breaking of $SO(10)$ in six dimensions by
the compactification of one dimension. 
Within 5D $N=1$ SUSY $G_{422}$ and $G_{51}$ models an extension with a
discrete
${\cal Z}$ symmetry is needed for a simultaneous solution of the 
problems {\bf (i)}-{\bf (vi)}.   
These models involve SM singlet right handed states which are necessary
for the breaking of the rank and obtaining the $G_{321}$ gauge group. In
combination
with the ${\cal Z}$ symmetry, these singlet states also play a crucial
role in
understanding of problems {\bf (i)}, {\bf (v)}, {\bf (vi)}. They are also
tied with lepton number violation and the generation of an intermediate
symmetry breaking scale.
 The $G_{422}$ model allows
to lower the unification scale not only down to intermediate scales,
but even
down to the multi TeV region. Differently, within $G_{51}$ the
unification scale is close to $\sim 10^{16}$~GeV.  
The models considered  have some peculiar 
phenomenological implications testable in the future.

The paper is organized as follows. In section 2 we present the main
construction principles of the models considered. In section 3 we write
the needed
one loop renormalization-group equations (RGE). Using them we
study gauge coupling unification within various models,  in the presence
of
KK states. Sections 4 and 5 are devoted to the 5D orbifold $N=1$ SUSY
$G_{321}$ and the $SU(5)$ models resp. 
In section 6 we discuss the issue of power law unification, within
orbifold GUT scenarios, and outline the
ways of its realization.
In sections 7 and 8
Pati-Salam $G_{422}$ and flipped $SU(5)\tm U(1)$ GUTs resp. are studied on
an
$S^{(1)}/Z_2\tm Z_2'$ orbifold. Finally discussions and conclusions are
presented in section 9. The paper contains an Appendix A, in which the
influence of some brane couplings on the gauge coupling running is
estimated.

\locsection{Construction principles of 5D SUSY \\
orbifold theories}

In this section we present our construction principles of 5D SUSY
theories. As we will see they are divided into  two categories: 
principles
which are related to the higher dimensionality and
others which deal with problems existing on the 4D level, 
after dimensional reduction.

{\bf $1^0$. 5D SUSY action}

We start the construction with a
5D $N=1$ SUSY theory. From the viewpoint of 4D (with coordinates $x$),
with the fifth coordinate $x_5\equiv y$ as a parameter, it is equivalent
to $N=2$ SUSY. $N=2$
supermultiplets can be expressed in terms of the usual 4D $N=1$
supermultiplets \cite{5dact}:
a gauge supermultiplet $V_{N=2}=(V,~\Si )$ contains the 4D $N=1$ vector
superfield $V$ and the chiral superfield $\Si $,
both in the adjoint representation of the gauge group
$G$ and  depending on the fifth coordinate. The 5D matter
superfield, in 4D language, is the
$N=2$ chiral supermultiplet ${\bf \Phi }_{N=2}=(\Phi,~\ov{\Phi })$,
where $\Phi$ is the $N=1$ chiral superfield and $\ov{\Phi }$ is it's
conjugate
-the so-called mirror (through out the paper the mirrors will be denoted by
an overline). So, if $\Phi $ is in some  irreducible representation
${\bf r}$ of $G$,
then $\ov{\Phi }$ will be in an antirepresentation ${\bf \bar r}$ 
of $G$.

Under gauge transformations one has
$$ 
e^V\to e^{\La }e^Ve^{\La^{+}}~,~~~~
\Si \to e^{\La }(\Si-\sq{2}\pl_5)e^{-\La }~,
$$
\beq 
\Phi \to e^{\La }\Phi~,~~~~~~~~\ov{\Phi } \to \ov{\Phi }e^{-\La }~,
\la{n2gaugetr}
\eeq
where $\La $ is a chiral superfield. The transformation of $\Si $ in 
(\ref{n2gaugetr}) reflects the 5D gauge invariance, since $\Si $ contains
the fifth component of the five dimensional gauge field \cite{highSUSY},
\cite{5dact}.
The 5D action can be written in terms of 4D $N=1$ superfields \cite{5dact}
and has
the form
\beq 
S^{(5)}=\int d^5x({\cal L}^{(5)}_V+{\cal L}^{(5)}_{\Phi })
\la{5dact},
\eeq  
where
$$ 
{\cal L}^{(5)}_V=\fr{1}{4g^2}\int d^2\te W^{\al }W_{\al }+{\rm h.c.}+
$$
\beq
\fr{1}{g^2}\int d^4\te \l (\sq{2}\pl_5V+\Si^{+})e^{-V}
(-\sq{2}\pl_5V+\Si )e^{V}+
\pl_5e^{-V}\pl_5e^{V}\r~,
\la{lagv}
\eeq 
\beq
{\cal L}^{(5)}_{\Phi }=\int d^4\te \l \Phi^{+}e^{-V}\Phi +
\ov {\Phi }e^{V}\ov {\Phi }^{+} \r+
\int d^2\te \ov{\Phi}\l M_{\Phi }+\pl_5-\fr{1}{\sq{2}}\Si \r \Phi+
{\rm h.c.}.
\la{lagf}
\eeq
Here $W_{\al }$ is the field strength supermultiplet, also in the adjoint
representation
of $G$
and built from $V$ ( $W_{\al }=-\fr{1}{4}\bar{D} \bar{D}D_{\al }V$). The
last term in (\ref{lagf}) contains the $F$-term of $\ov{\Phi }\pl_5 \Phi
$,
which is crucial for 5D Lorentz invariance: for a bosonic component
$\Phi_s$ of the $\Phi $ superfield it produces the term $|\pl_5 \Phi_s|^2$
which
together with $|\pl_{\mu }\Phi_s|^2$ [coming from the first coupling in
(\ref{lagf})] is 5D Lorentz invariant. The same happens for the fermionic
components.  The $\pl_5-\fr{1}{\sq{2}}\Si $
combination is crucial for the 5D gauge invariance under (\ref{n2gaugetr}).

There are two supersymmetries in ${\cal L}^{(5)}$, the obvious
4D $N=1$ SUSY and one related by a global $SU(2)_R$ symmetry
to the former one \cite{highSUSY}. Thus the SUSY transformation parameters
as well as the scalar components of $(\Phi,\ov{\Phi })$ and the two
spinors  
$(\lambda,\lambda')$ in $W_{\al }$
and $\Si $ form doublets under this $SU(2)_R$. The fermionic
components of $(\Phi,\ov{\Phi })$ and the bosonic components
of $W_{\al }$ and $\Si $ are $SU(2)_R$ singlets. The $N=1$ SUSY
theory in 5D has the advantage that there is no free superpotential.
The action is completely fixed except for the $M_{\Phi }$ term
in (\ref{lagf}) which
in some cases might be forbidden by orbifold parities (see below).
The $M_{\Phi }$ only connects fields with their mirrors.

{\bf $2^0$. Compactification and orbifold symmetries}

Since we have one extra dimension, it is important somehow to reduce the
theory to the 4D one. One can start from a 
${\cal M}^{(4)}\otimes S^{(1)}$ theory, where ${\cal M}^{(4)}$ is the four
dimensional Minkowski space-time and $S^{(1)}$ a compact
circle. Equivalently, one can consider the fifth dimension as an
infinite
$R^{(1)}$ line and impose some periodicity 
$y\sim y+2\pi R$, where $R$ is the radius of the circle corresponding to
the
characteristic compactification scale $\mu_0\simeq 1/R$.
So, the theory in the fifth dimension is defined on a
interval $L'=[0,~2\pi R]$ or
equivalently on $L=[-\pi R,~\pi R]$. 
On the interval $L$ one  can introduce discrete symmetries, and $Z_2$ is
the simplest one

\beq
Z_2~:~~~~~y\to -y~,
\la{Z2orbsym}
\eeq
which folds the circle. The theory is then built on an $S^{(1)}/Z_2$
orbifold.
Under (\ref{Z2orbsym}) all introduced fields $\phi $ should have definite
parity transformation properties $\phi (x, y) \to P\phi (x, y) $, such
that the 5D Lagrangian
(\ref{lagv}), (\ref{lagf}) is invariant ($\phi $ designates all  
gauge and matter supermultiplets we have). $P=\pm 1$ and the mode
expansions of states $\phi_{+}$ and $\phi_{-}$ with positive and negative
parities resp. have the form

\beq
\phi_{+}(x, y)=\sum_{n=0}^{\infty }\phi^{(n)}_{+}(x)
\cos \fr{ny}{R}~,~~~
\phi_{-}(x, y)=\sum_{n=1}^{\infty }\phi^{(n)}_{-}(x)
\sin \fr{ny}{R}~,
\la{modeZ2}
\eeq
$\phi^{(n)}_{+}$ and $\phi^{(n)}_{-}$ are Kaluza-Klein (KK) states.
As we see, $\phi_{-}(x, y)$ does not contain a zero mode. Massive KK modes
have masses $m_{n}^{KK}=n/R=n\mu_0$. We have two fixed points $y=0$ and
$y=\pi R$. With the help of the $Z_2$ orbifold parity it is possible to
project
out some states (assigning them negative parities) and to achieve
the breaking
of supersymmetries and gauge symmetries. 
If we wish to break the gauge group $G$ down to its subgroup $H$, 
gauge
fields $V(G/H)$ should have negative parities, while the parities of
fragments
$V(H)$ are positive.
From (\ref{lagv}), it is
clear that in this case $P[\Si (H)]=-1$ and $P[\Si (G/H)]=+1$
(because $y$ changes sign under $Z_2$). Also, it follows from 
(\ref{lagf}) that mirrors must have opposite parities. Because of all
this,
together with the gauge symmetry, half of the SUSY is broken and at the
fixed points we have a $4D$ $N=1$ SUSY theory with a reduced gauge group.
But we are also left with the additional zero mode states of $\Si (G/H)$.
In order to avoid them, the orbifold symmetry can be extended to
$Z_2\tm Z_2'$ \cite{kawa}:
by additional folding of the half circle
\beq
Z_2~:~~y\to -y~,~~~~~~Z_2'~:~~y'\to -y'~,
\la{orbZs}
\eeq
where $y'=y+\fr{\pi R}{2}$, one can ascribe negative $Z_2'$ parity to 
$\Si (G/H)$ and '$Z_2'$ charge' for $V(G/H)$. Now the theory is defined on
an 
$S^{(1)}/Z_2\tm Z_2'$ orbifold and at the $y=0$ fixed point (identified
with
our 4D world $3$-brane) we have a 4D $N=1$ SUSY theory with gauge
group $H$. No additional fragments of $\Si $ with zero mode wave
functions emerge. Of course, also in this case mirrors 
should
have opposite $Z_2'$ parities.
In the next sections we will
demonstrate transparently with concrete examples how this procedure is
realized.
Each state has a definite $Z_2\tm Z_2'$
parity $(P, P')$ parity $\sim (\pm ,~\pm )$.
Therefore,
under the transformations (\ref{orbZs}):
\beq
\phi \to P\phi ~,~~~~~~~\phi \to P'\phi ~.
\la{transpar}
\eeq
Depending on the $(P, P')$ parity, there are four
possible mode expansions $\phi_{\pm  \pm}$
$$
\phi_{++}(x, y)=\sum_{n=0}^{\infty }\phi^{(2n)}_{++}(x)\cos \fr{2ny}{R}~,
$$
$$
\phi_{+-}(x, y)=\sum_{n=0}^{\infty }
\phi^{(2n+1)}_{+-}(x)\cos \fr{(2n+1)y}{R}~,
$$
$$
\phi_{-+}(x, y)=\sum_{n=0}^{\infty }
\phi^{(2n+1)}_{-+}(x)\sin \fr{(2n+1)y}{R}~,
$$
\beq
\phi_{--}(x, y)=\sum_{n=0}^{\infty }
\phi^{(2n+2)}_{--}(x)\sin \fr{(2n+2)y}{R}~.
\la{modexp}
\eeq
Consequently, the masses of the appropriate Kaluza-Klein (KK) modes of
$\phi^{(2n)}_{++}$, $\phi^{(2n+1)}_{+-}$, $\phi^{(2n+1)}_{-+}(x)$ and
$\phi^{(2n+2)}_{--}$
will be $\fr{2n}{R}$, $\fr{2n+1}{R}$, $\fr{2n+1}{R}$,
and $\fr{2n+2}{R}$, resp.
Only the $\phi_{++}$ states contain massless zero modes. States
with
other parities are massive.
We emphasize again that, if we introduce states in the bulk and ascribe
to them some parity $(p, p')$,  
the mirror must carry $(-p, -p')$ parity. In this way we have 5D Lorentz 
invariance.  
This is quite different when a state is fully restricted to the
brane and does not have KK excitations (as possibly chiral
matter in some cases which we will consider below)
\footnote{
The latter scenario has not much to do with the orbifold symmetries which
we
consider here. 
It can be realized if states are confined on intersecting 
branes \cite{branint}. 
}. If we want a
state (introduced in the bulk) to have a zero
mode component, we should assign to it (+, +) parity. For all other
parity choices, the states have only massive KK excitations.

{\bf $3^0$. Construction of the 4D theory on a brane, 

additional discrete symmetries and extensions}

As we have already mentioned, 5D SUSY does not allow to have a
superpotential which leads to Yukawa
couplings. This enforces  brane couplings in
order to build a realistic phenomenology. Couplings at the  $y=0$
fixed point\footnote{We are selecting the fixed point
which is more suitable for realistic model building as
our 4D world.}
\beq
{\cal L}'=\int d y\de (y) W^{(4)}(x, y)~,
\la{branesup}
\eeq
possess 4D $N=1$ supersymmetry and involve fields with zero mode wave
functions. $W^{(4)}$ includes Yukawa couplings which are responsible for
the
generation of fermion masses. The couplings in (\ref{branesup}) do not
violate the
higher supersymmetries and gauge symmetries of the 5D bulk. The reason for
this
is that 
the wave functions of generators which transform zero mode states to 
states with negative
orbifold parities vanish in the 4D fixed point. In this way the whole
theory is self consistent.

As we have already mentioned in the introduction, the orbifold
constructions
have big advantages in resolving various puzzles connected with
GUTs. However,  the  problems 
{\bf (i)}, {\bf (v)}-{\bf (vii)} (mentioned in the introduction) 
still remain at the 4D
level, and need to
be tackled. Amongst them the most urgent ones are 
baryon number conservation and the $\mu $ problem. 
Furthermore,  problems emerging from matter parity
violating operators  should be avoided and the neutrino deficits must be
explained. We consider these problems to be severe enough 
to motivate us to think
about some
reasonable extension of the considered scenario.
Starting with the $\mu $ problem, for its solution we introduce
an additional
discrete symmetry ${\cal Z}$ and prescribe transformation
properties to 
$h_u$, $h_d$ in such a way as to forbid a direct $\mu $ term.  
We also introduce singlets 
${\cal S}$, $\ov{\cal S}$ which have VEVs $\ll M $ (the cutoff scale).
Through 
$\l \fr{\ov{\cal S}{\cal S}}{M^2} \r^{n}h_uh_u$ type couplings with a  
proper
choice of $n$ we
obtain a $\mu $ term of the desired magnitude \cite{mugen1}.
In section 4, we explicitly demonstrate how the generation of 
$\ov{\cal S}$, ${\cal S}$
VEVs and the $\mu $ term suppression are realized. The MSSM and the 
minimal SUSY $SU(5)$
require ${\cal S}$, $\ov{\cal S}$ singlets, while the models
$SU(4)_c\tm SU(2)_L\tm SU(2)_R$ and flipped $SU(5)\tm U(1)$ automatically
involve scalars  being singlets of the MSSM (see sections 7, 8).

In the MSSM and SUSY GUTs, usually a $Z_2$ $R$-parity is assumed, which
distinguishes matter and scalar superfields and avoids baryon number and
large lepton number violation. In our approach, for the same purpose we  
use the ${\cal Z}$ symmetry, which avoids all baryon number violating
couplings which also violate $R$ parity.
With help of the introduced ${\cal Z}$ symmetry we also avoid
$d=5$ and $d=6$ baryon number violating Planck scale operators, which are
otherwise allowed on the 4D level, causing unacceptably rapid nucleon
decay ($d=6$
operators become dangerous if we are dealing with low or
intermediate scale theories). So, from this point of view, the extension
with a
discrete
${\cal Z}$ symmetry turns out to be very efficient
\cite{ZvsBviol}
\footnote{For the same purposes discrete, continuous ${\cal R}$ 
\cite{RvsBviol} and anomalous gauge ${\cal U}(1)$ \cite{u1vsBviol}
symmetries have been used. In \cite{gaugeB}  models
with gauged baryon number were suggested.}.

As far as the lepton
number violating
couplings are concerned it is well known that the MSSM and the minimal
SUSY
$SU(5)$ do not give sufficiently large neutrino masses and that, for
accommodation of atmospheric and solar neutrino data, some extensions 
are needed. In our constructions we admit some lepton number violating
couplings (which usually are absent due to $R$ parity) and due to proper
suppression (with the help of the ${\cal Z}$ symmetry) they give desirable
value(s) for the
neutrino masses. We will discuss this issue in more detail through the
sections 4, 5 and 8. 

Concluding this section we point out that,  when using ${\cal
Z}$ symmetry, one should make the corresponding charge assignments to
the matter and
scalar supermultiplets in such a way that the terms in (\ref{lagf}),
allowed
by orbifold symmetries, are invariant also under ${\cal Z}$
\footnote{This requirement does not apply for matter states which do not 
live in the bulk, but are introduced only at a fixed point brane.}.
This means that mirrors must have opposite '${\cal Z}$ charges' and if the
considered scenario is a GUT, the states coming from one unified multiplet
should have the same transformation properties under the ${\cal Z}$
symmetry.

\locsection{Renormalization-group equations}

In this section we will present general expressions for the solutions of
the one
loop renormalization-group equations (RGE) in the presence of KK
excitations
corresponding to one extra space like dimension, which will be needed
to estimate  gauge coupling unification in different scenarios. At
energy scales below the compactification scale $\mu_0=1/R$ the one loop
running of the gauge couplings $\al_i$ has logarithmic form \cite{RGE}
\beq
\al_i^{-1}(\mu_{\rho+1} )=\al_i^{-1}(\mu_{\rho })-
\fr{b_i^{\rho }}{2\pi }\ln
\fr{\mu_{\rho+1}}{\mu_{\rho }}~.
\la{RGElog}
\eeq
{}For the standard model
the gauge groups labeled by 
$i=1, 2, 3$ correspond to $U(1)_Y$, $SU(2)_L$, $SU(3)_c$
resp. Without intermediate scales and additional states, 
the $b_i^{\rho }\equiv b_i$ factors will be just those corresponding to
the
states of the SM or the MSSM (depending on whether the theory we 
are studying is supersymmetric or not). Assume that up to a certain
mass
scale $M_I$ we
have the $SU(3)_c\times SU(2)_L\times U(1)_Y\equiv G_{321}$ gauge group
with the
minimal content of SM/MSSM. Then the couplings at $M_I$ are
\beq
\al_a^{-1}(M_I)=\al_a^{-1}(M_Z)-
\fr{b_a}{2\pi }\ln \fr{M_I}{M_Z}~.
\la{RGE321}
\eeq
Labeling couplings in (\ref{RGE321}) by $a$ we emphasize that we are
dealing with $G_{321}$ gauge couplings. Above the scale $M_I$ the gauge
group can be different and consequently runnings should be studied
according to the existing gauge group and the corresponding  
states. Couplings at
different mass
scale regions must be matched at the intermediate scale(s) $M_I$. So, we
will run couplings up to the unification scale $M_G$, which we treat as
the
cutoff scale of a theory. Since we are considering theories with one
compact dimension, above the scale $\mu_0$ we should include the effects
of KK modes.
In the concrete models considered below, at an intermediate
scale
$M_I$, two gauge groups [either two $U(1)$s or $SU(2)$ and $U(1)$]
are reduced to the $U(1)_Y$. We have the boundary/matching
condition
\beq
\al_1^{-1}(M_I)=\sin^2 \te \cdot \al_{G_1}^{-1}(M_I)+
\cos^2 \te \cdot \al_{G_2}^{-1}(M_I)~,
\la{match}
\eeq
where $\tan \te $ is a group-theoretical factor determined from the
pattern of $U(1)_Y$ gauge group embedding in a product group 
$G_1\times G_2$. $\al_{G_1}$, $\al_{G_2}$ are the couplings of the gauge
groups
$G_1$, $G_2$ and above $M_I$ we will have equations of
the (\ref{RGElog}) type 
for them. However, we can also write RGE for the
combinations (\ref{match}) in 
(\ref{RGElog}) form, where the role of $b_1$ is now to be played by 
a superposition of 
$b_{G_1}$ and $b_{G_2}$, similar to (\ref{match})
\beq
b_1^{M_I}=\sin^2 \te \cdot b_{G_1}+\cos^2 \te \cdot b_{G_2}~.
\la{BsupG1G2}
\eeq
Taking all this into account, we will have
\beq
\al_a^{-1}(M_G)=\al_a^{-1}(M_Z)-\fr{b_a}{2\pi }\ln \fr{M_I}{M_Z}+\De_a~,
\la{alGa}
\eeq
with
\beq
\De_a=\De_a^0+\De_a^{KK}~,
\la{Delta}
\eeq
where
\beq
\De_a^0=-\fr{(b_a^{M_I})_{\al }}{2\pi }\ln \fr{M_G}{(M_I)_{\al }}~
\la{Delog}
\eeq
includes contributions from all existing zero mode states $\al $ with mass
$(M_I)_{\al }$. $\De_a^{KK}$ comes from the contributions of KK states. In
the case that their masses are $n\mu_0$, we  have 
\beq
\De_a^{KK}=-\fr{\hat{b}_a}{2\pi }S~,~~~~
S=\sum_{n=1}^{N_0}\ln \fr{M_G}{n\mu_0}~,
\la{DeKKZ}
\eeq
where $\hat{b}_a$ is a common factor
of the given KK states and $N_0$ stands for the maximal number
of
KK states which lie below $M_G$.

For models with $Z_2\times Z_2'$ orbifold parities, $\De_a^{KK}$ will
have the form
\beq
\De_a^{KK}=-\fr{\ga_a}{2\pi }S_1-\fr{\de_a}{2\pi }S_2~,
\la{DeKKZZ1}
\eeq
where $S_1$ and $S_2$ include contributions
from KK states with masses $(2n+2)\mu_0$ and $(2n+1)\mu_0$ resp.:
\beq
S_1=\sum_{n=0}^{N} \ln \fr{M_G}{(2n+2)\mu_0}~,~~~~
S_2=\sum_{n=0}^{N'} \ln \fr{M_G}{(2n+1)\mu_0}~.
\la{SKK}
\eeq
In (\ref{SKK}), $N$ and $N'$ are the maximal numbers of appropriate KK
states
which lie below $M_G$, i.e.
\beq
(2N+2)\mu_0\stackrel{<}{_\sim }M_G~,~~~~~
(2N'+1)\mu_0\stackrel{<}{_\sim }M_G~.
\la{maxNs}
\eeq
KK states with masses larger than $M_G$ are irrelevant.  
For a given
$M_G/\mu_0$ the $N$ and $N'$ can be calculated from (\ref{maxNs}).
Let us note, that $(b_1^{M_I})_{\al }$, $\ga_1$ and $\de_1$ will be
expressed by similar superpositions as $b_1^{M_I}$ in (\ref{BsupG1G2}),
\beq
\ga_1=\sin^2 \te \cdot \ga_{G_1}+\cos^2 \te \cdot \ga_{G_2}~,~~~
\de_1=\sin^2 \te \cdot \de_{G_1}+\cos^2 \te \cdot \de_{G_2}~.
\la{KKsupG1G2}
\eeq
If at a scale $M_G$ we impose the condition of gauge coupling unification
\beq
\al_1(M_G)=\al_2(M_G)=\al_3(M_G)\equiv \al_G~,
\la{coupuni}
\eeq
then from (\ref{alGa}), eliminating $\al_G$ and $\ln M_I/M_Z$, we find for
the strong
coupling at the $M_Z$ scale 
\beq
\al_3^{-1}=\fr{b_1-b_3}{b_1-b_2}\al_2^{-1}-
\fr{b_2-b_3}{b_1-b_2}\al_1^{-1}+
\fr{b_1-b_3}{b_1-b_2}\De_2-\fr{b_2-b_3}{b_1-b_2}\De_1-\De_3~,
\la{als}
\eeq
where $\al_a$ in (\ref{als}) stands for $\al_a(M_Z)$.
Also, from (\ref{alGa}) one can obtain 
\beq
\ln \fr{M_I}{M_Z}=\fr{2\pi }{b_1-b_2}(\al_1^{-1}-\al_2^{-1})+
\fr{2\pi }{b_1-b_2}(\De_1-\De_2)~,
\la{scale}
\eeq
and finally the value of the unified gauge coupling
\beq
\al_G^{-1}=\al_2^{-1}-\fr{b_2}{2\pi }\ln \fr{M_I}{M_Z}+\De_2~.
\la{alG}
\eeq

{}For a given model, the values of $\De_a$ can be  fixed
[according to (\ref{Delta}), (\ref{Delog}), (\ref{DeKKZ}) or
(\ref{DeKKZZ1})] and from (\ref{als}) one can calculate $\al_3$. The
contribution from the $\De_a$s should not be too large, such that
the experimental value
\cite{pdg} $\al_3(M_Z)=0.119\pm 0.002$ is obtained. If the contributions
from $\De_a$ in (\ref{scale}) are negative and large, one can obtain a
(relatively) low $M_I$ scale and consequently
small
$\mu_0$, $M_G$. When constructing models, we should keep in the mind that
the gauge
couplings must remain in the perturbative regime until they reach the
unification point. For this it is enough to require a perturbative value
for
$\al_G$, calculated from (\ref{alG}). 

In the following, equations (\ref{als})-(\ref{alG}) will be used 
to estimate
the status of gauge coupling unification in various scenarios.
Of course, taking into account various threshold corrections (from weak
and GUT scales or from some brane localized operators), these equations
will have additional entries. The relevance of such contributions
will be commented below.

\locsection{5D SUSY $SU(3)_c\tm SU(2)_L\tm U(1)_Y$ on $S^{(1)}/Z_2$
orbifold}
  
Consider a 5D $N=1$ SUSY $SU(3)_c\tm SU(2)_L\tm U(1)_Y\equiv G_{321}$
theory. Since we do not have to break the gauge group it is enough to
introduce
only one $Z_2$ orbifold parity, i.e. the theory is defined on
an $S^{(1)}/Z_2$
orbifold. According to the discussions of section 2, in this way we can
break
half of the supersymmetries. 
The field content, their orbifold parities and $Y$ hypercharges 
are given in Table \ref{t:2}. We use the $SU(5)$
normalization 
\beq
Y=\fr{1}{\sq{60}}(2,~2,~2,~-3,~-3)~.
\la{hipcharge}
\eeq

At the $y=0$ fixed point we are left with the SUSY $G_{321}$ gauge theory
with zero mode states $q$, $l$, ${u^c}$, ${d^c}$,
${e^c}$, $h_u$, ${h_d}$, which is just the content of the MSSM.

{\bf Fixed $y=0$ point brane couplings 

and some phenomenology }

In order to build a realistic theory we  write brane
couplings of the (\ref{branesup}) type. The
4D Yukawa superpotential, responsible for the generation of up-down quark
and
charged lepton masses, has the form (neglecting coupling constants)

\beq
W_Y=q{u^c}h_u+q{d^c}{h_d}+l{e^c}{h_d}~.
\la{yukMSSM}
\eeq 

According to  part {\bf $3^0$}  of section 2, to
resolve various problems, it is useful to introduce a ${\cal Z}$ discrete
symmetry.
With the symmetry transformation
\beq
h_u{h_d}\to e^{{\rm i}\fr{2\pi }{n}}h_u{h_d}~, 
\la{mutermMSSM}
\eeq
the $Mh_u{h_d}$ coupling is forbidden. Introducing singlet states ${\cal
S}$,
$\ov{\cal S}$ \footnote{${\cal S}$, $\ov{\cal S}$ states can be
introduced in the bulk. In this case, on the 5D level they are accompanied
by the
appropriate mirrors with opposite orbifold parities. For us the
4D superpotential couplings are important in which only the zero modes of
${\cal S}$, $\ov{\cal S}$ participate.} with
the transformation

\beq
\ov{\cal S}{\cal S}\to e^{{\rm i}\fr{2\pi }{n}}\ov{\cal S}{\cal S}~,
\la{Ssym}
\eeq
we get that the relevant coupling will be
\beq
W_{\mu }=M\l \fr{\ov{\cal S}{\cal S}}{M^2}\r^{n-1}h_u{h_d}~.
\la{supmuMSSM}
\eeq
%
%
\begin{table} \caption{Hypercharges and $Z_2$ parities of the states
within 5D $N=1$ SUSY $G_{321}$.}
 
\label{t:2} $$\begin{array}{|c|c|c|}
 
\hline 
{\rm Superfields } &\sqrt{60}\cdot Y & Z_2 \\
 
\hline \hline
 
V_c, ~V_{SU(2)_L}, ~V_Y & 0 & +\\
 
\hline

\Sigma_c,~\Sigma_{SU(2)_L},~\Sigma_Y & 0 & -\\ 

\hline \hline

{\bf q}_{N=2}=( q,~\ov{q}) &( -1,~1) &(+,~-) \\ 

\hline 

{\bf l}_{N=2}=( l,~\ov{l}) & (3,~-3) &(+,~-)  \\ 

\hline 

{\bf u^c}_{N=2}=( u^c,~\ov{u}^c) & (4,~-4) &(+,~-)\\ 

\hline 

{\bf d^c}_{N=2}=( d^c,~\ov{d}^c) & (-2,~2) &(+,~-)\\

\hline 

{\bf e^c}_{N=2}=( e^c,~\ov{e}^c) &( -6,~6) &(+,~-)\\

\hline\hline

{\bf h^u}_{N=2}=(h_u~,~\ov{h}_u) & (-3,~3) & (+,~-) \\

\hline

{\bf h^d}_{N=2}=(h_d~,~\ov{h}_d) & (3,~-3) & (+,~-)\\

\hline

\end{array}$$
 
\end{table}
%
%
Due to the transformations of $h_uh_d$ and $\ov{{\cal S}}{\cal S}$, 
${\cal Z}$ acts as a $Z_n$ symmetry.
If ${\cal S}$, $\ov{{\cal S}}$ develop VEVs such that 
$\lan {\cal S}\ran, \lan \ov{{\cal S}}\ran \ll M $, by an adequate
choice of $n$ one can obtain a properly suppressed $\mu $ term.
The lowest
superpotential coupling for ${\cal S}$, $\ov{{\cal S}}$ is
\beq
W_{\cal S}=\lam M^3\l \fr{\ov{{\cal S}}{\cal S}}{M^2}\r^n~,
\la{supS}
\eeq
and in the unbroken SUSY limit the conditions 
$F_{\cal S}=F_{\ov{\cal S}}=0$ give
$\lan {\cal S}\ran =\lan \ov{\cal S}\ran =0$. 
After SUSY breaking, soft
SUSY breaking terms
should be involved. The relevant soft terms concerning 
${\cal S}$, $\ov{{\cal S}}$ are
\beq
V_{soft}({\cal S})=m_1^2 |{\cal S}|^2+m_2^2|\ov{\cal S}|^2 
+m_3A(W_{\cal S}+W^*_{\cal S})~,
\la{softSpot}
\eeq
where $m_1$, $m_2$, $m_3$ are all of order of the SUSY scale $m$ and $A$
is a dimensionless constant.
With (\ref{supS}), (\ref{softSpot}) one can write the total potential
for
${\cal S}$ as
\beq
V({\cal S})=|F_{\cal S}|^2+|F_{\ov{\cal S}}|^2+V_{soft}({\cal S})~.
\la{potS}
\eeq
Minimization of (\ref{potS}) leads to
a non zero solution for $\lan {\cal S} \ran$, $\lan \ov{{\cal S}}\ran $
\beq
\lan {\cal S} \ran \sim \lan \ov{{\cal S}}\ran 
\sim M\l \fr{m}{M}\r^{\fr{1}{2n-2}}~.
\la{solS}
\eeq 
Substituting (\ref{solS}) into  (\ref{supmuMSSM}), we obtain for the 
$\mu $ term
\beq
\mu \simeq M\l \fr{\lan {\cal S} \ran}{M}\r^{2n-2}\sim m~.
\la{valmuterm}
\eeq
As we see the $Z_n$ symmetry gives a natural generation of the $\mu $ 
term (which is independent of $n$!)
with the required magnitude. 
The relevant feature is that the $\lan {\cal S} \ran $,
$\lan \ov{{\cal S}} \ran $ in (\ref{solS}) are
expressed through the interplay of the two scales $m$ and $M$
\cite{scalegen}.
For $m=1$~TeV, $M=M_P=2.4\cdot 10^{18}$~GeV (reduced Planck mass) and
$n=9$ one has $\lan {\cal S} \ran /M\sim 1/10$. For lower values of the
fundamental scale $M$, the desired gap between $\lan {\cal S} \ran $ and
$M$
can be obtained by a proper choice of $n$. For example, for the same
value of
$m$, $\lan {\cal S} \ran/M  $ and $M\sim 10^{13}$~GeV, 
we need $n=6$, while for $M\sim 100$~TeV no large
suppression is required and one can take $n=2$.

The introduced discrete symmetry is crucial for avoiding $d=5$ baryon
number
violation and also unacceptably large $R$-parity violating operators.
We note however that amongst the latter operators
there are lepton number violating couplings which, being properly
suppressed, could generate neutrino masses of the 
needed magnitude \cite{penRpar}-\cite{symRsupr3}.
Together with the Yukawa couplings in (\ref{yukMSSM}), we can
therefore include the lepton
number
violating  coupling 
\beq
M\l \fr{\ov{\cal S}{\cal S}}{M^2}\r^{n-1}
\l \fr{\cal S}{M} \r^kh_ul~,
\la{bilinS}
\eeq
which after substituting the appropriate VEVs [see (\ref{solS}),
(\ref{valmuterm})] leads to the bi-linear operator
\beq
\mu_l h_ul~,~~~\mu_l\sim \l \fr{\lan {\cal S}\ran }{M}\r^k\mu ~.
\la{bilinopS}
\eeq
Due to this operator the sneutrino field can gain a VEV of the order 
$\lan \tl{\nu }\ran\sim \fr{\mu_l}{\mu }{\cal O}(100~{\rm GeV})
\equiv \sin \xi {\cal O}(100~{\rm GeV})$. The 
latter produces a neutrino-neutralino mixing which leads to a  neutrino
mass
through the see-saw type mechanism
\cite{bilRpar} 
\beq
m_{\nu }\sim {\cal O}(100~{\rm GeV})\sin^2 \xi~,
\la{nubilinS}
\eeq
where $\sin \xi \sim \mu_l/\mu$ (assuming that there is no alignment
between the
superpotential and the soft SUSY breaking couplings). 
To have a neutrino mass $\stackrel{<}{_\sim }1$~eV, in (\ref{nubilinS}) we
need $\sin \xi\stackrel{<}{_\sim }3\cdot 10^{-6}$.
With a $\mu $-term
$\sim m$ and $k=6$, $5$, $\lan {\cal S}/M\ran 
\sim 1/10$ - $1/15$, 
from
(\ref{bilinopS}) we
have
$\sin \xi \sim 10^{-6}$ which gives $m_{\nu_{\tau }}\sim 0.1$~eV,
indeed the order of magnitude needed for explaining the atmospheric
neutrino anomaly.

The  phases of $h_u$, ${h_d}$, ${\cal S}$, $\ov{{\cal S}}$
were not fixed by the couplings (\ref{supmuMSSM}) , (\ref{supS}). 
The  couplings given above determine the transformation properties
of the different 
states under the ${\cal Z}$ symmetry
$\phi_i\to e^{{\rm i}\al (\phi_i)}\phi_i $ [$\al (\phi_i)$ is 
the phase of state $\phi_i$, 
and its mirror $\ov{\phi}_i$ has opposite phase).
Due to the couplings in (\ref{yukMSSM}), (\ref{supmuMSSM}), 
(\ref{supS}), (\ref{bilinS}) we have
$$
\al (\ov{{\cal S}})=\al -\al ({\cal S }) ~,~~~
\al ({h_d})=\al -\al (h_u)~,~~  
$$
$$
\al ({d^c})=-\al (q)+\al (h_u)-\al ~,~~ 
\al (l)=-\al (h_u)-k\al ({\cal S})+\al  ~,
$$
\beq
\al ({u^c})=-\al (q)-\al (h_u)~,~~~
\al ({e^c})=2\al (h_u)+k\al ({\cal S})-2\al ~,~~
\al =\fr{2\pi }{n}~,
\la{ZphasesMSSM}
\eeq
where $\al (q) $, $\al (h_u)$, $\al ({\cal S})$ are undetermined. 
Other allowed $R$ parity breaking operators also violating the lepton
number are
\beq
\l \fr{\cal S}{M}\r^{k}q{d^c}l~,~~~~
\l \fr{\cal S}{M}\r^{k}{e^c}ll~.
\la{trilinS}
\eeq
After substituting VEV of ${\cal S}$ they lead to the couplings
\beq
\lam qd^cl~,~~~\lam' e^cll~,~~~
\lam \sim \lam' \sim \l \fr{\lan {\cal S}\ran }{M}\r^k~.
\la{trilinS1}
\eeq
These couplings induce neutrino masses at one loop, with the dominant
contribution given by the $b^c$ state inside the loop, 
\beq
{m_{\nu }}'\propto \fr{\lam^2}{8\pi^2}\fr{m_b^2m}{m_{\tl b}^2}~,
\la{numastrilin}
\eeq
which for $k=4-6$, $\lan {\cal S}\ran/M\sim 1/10$, $m=1$~TeV, 
$m_{\tl b}=300$~GeV is evaluated as
$m_{\nu }' \sim (10^{-2}-3\cdot 10^{-6})$~eV, 
to explain the solar neutrino puzzle either through MSW
(by large or small angle, depending on which
mixing scenario is realized for the fermion sector) or large angle
vacuum oscillations (LAVO). 
This way of neutrino mass generation through properly suppressed $R$
parity violating operators \cite{penRpar}-\cite{symRsupr3} looks
attractive since it does not
require the
introduction of right handed neutrinos. However, additional symmetries
(in this case ${\cal Z}$) are crucial \cite{symRsupr1},
\cite{symRsupr3} for obtaining
properly suppressed neutrino masses.

With the assignments (\ref{ZphasesMSSM}) and taking $\al (q)=\al/2 $, 
$\al (h_u)=\al $, $\al ({\cal S})=\al /3$ the
discrete symmetry introduced would be $Z_{6n}$. With the phases presented,
the
${\cal S}^{3n}$ and other higher order terms are allowed, but along the
(\ref{solS}) solution they are strongly suppressed in comparison to terms
in
(\ref{supS}), (\ref{softSpot}). Therefore, the analyzes above stay valid. 
One can also verify that
for any integer $k$ the baryon number violating $d=4$ operator
$u^cd^cd^c$ is forbidden.
Also, the baryon number violating $d=5$ operators
\beq
\fr{1}{M}qqql~,~~~\fr{1}{M}{u^c}{u^c}{d^c}{e^c}~,~~~
\fr{1}{M}qqqh_d
\la{d5MSSM}
\eeq
are not allowed. 
There are also $d=6$ baryon number violating $D$-term operators
\beq
\fr{1}{M^2}\left [ qqu^{c+}e^{c+}\right ]_D~,~~
\fr{1}{M^2}\left [ qlu^{c+}d^{c+}\right ]_D~,~~
\fr{1}{M^2}\left [ qh_du^{c+}d^{c+}\right ]_D~,
\la{d6MSSM}
\eeq
which for low values of $M$ can become important and induce nucleon decay.
It is easy to check that they are also forbidden by the ${\cal Z}=Z_{6n}$
symmetry.

{\bf Unstable LSP}

With the presence of $R$ parity violating couplings, the LSP - the
lightest
neutralino $\chi $ - is an unstable particle. In the scenario considered,
the
LSP three body decays mostly proceed due to the bi-linear (\ref{bilinopS}) 
coupling, and the LSP lifetime is

\beq
\tau^{-1}_{\chi }=\mu_l^2Z_{\chi \tl{H}}^2
\l \fr{1}{4}+\sin^2 \te_W+\fr{4}{3}\sin^4 \te_W \r
\fr{G_F^2m_{\chi }^3}{192\pi^3}~.
\la{LSPtime}
\eeq 
{}For the value $\mu_l\sim 10^{-6}\mu \sim 10^{-3}$~GeV (dictated from
the atmospheric neutrino scale) we have 
$\tau_{\chi }\sim 10^{-20}$~sec. 
Therefore the LSP would be cosmologically irrelevant and some other
candidate
for cold dark matter should be found.

\subsection{Gauge coupling unification in 5D SUSY $G_{321}$}

Below the compactification scale $\mu_0$ the field content is just that of
the MSSM and the
corresponding $b$ factors are
\beq
(b_1,~b_2,~b_3)=\l \frac{33}{5},~1,~-3\r~.
\la{Bmssm}
\eeq
Above the $\mu_0$ scale the KK states enter into the
renormalization. Having KK excitations for all gauge and scalar
superfields and also for $\eta $ families of bulk matter, the $\hat{b}$
factors,
corresponding to the power law running (\ref{DeKKZ}), are

\begin{equation}
                   (\hat{b}_1,~\hat{b}_2, ~\hat{b}_3)=
\l \frac{6}{5},-2,-6\r+4\eta\,(1,~1,~1)~.
\la{KKMSSM}
\end{equation}
{}From (\ref{als})-(\ref{alG}), taking into account 
(\ref{DeKKZ}), (\ref{Bmssm}), (\ref{KKMSSM}), we obtain
\begin{equation}\label{eq:A}
           \alpha_3^{-1}= \frac{12}{7}\alpha_2^{-1}-\frac{5}{7}
\alpha_1^{-1} -\frac{6}{7\pi}S,
\end{equation}
\begin{equation}\label{eq:B}
           \ln{\frac{M_G}{M_Z}}= \frac{5\pi}{14}(\alpha_1^{-1} - 
\alpha_2^{-1})-\frac{4}{7}S,
\end{equation}
\begin{equation}\label{eq:C}
            \alpha_G^{-1}=\alpha_2^{-1}-\frac{1}{2\pi}
\ln{\frac{M_G}{M_Z}}+\frac{1-2\eta}{\pi}S,
\end{equation}
where $M_I=M_G$ was taken
since there is no intermediate scale .  
$\alpha_1^{-1}$, $\alpha_2^{-1}$ are known with high precision and 
$\alpha_3^{-1}$ with some precision and thus, eq.(\ref{eq:A}) sets a
constraint to 
the values of $S$. With $\alpha_1^{-1}=59$, $\alpha_2^{-1}=29.6$ and 
$0.117 \le \alpha_3 \le 0.121$ we get
\begin{equation}
             0.19\le S \le 1.23~.
\la{sazS}
\end{equation}
{}From eq.(\ref{eq:B}) we see that the \emph{unification} scale is 
also constrained:
\begin{equation}
             M_G \simeq (1 - 2)\cdot 10^{16}~{\rm GeV}.
\la{sazMG}
\end{equation}
From the definition
(\ref{DeKKZ})  of $S$ and (\ref{sazS}), 
(\ref{sazMG}) a constraint to the
values of $N_0$, 
the number of KK levels, arises.
%
It is not difficult to see that $N_0=1,~2$ are the only values of $N_0$ 
allowed for $0.19\le S \le 1.23$. Two examples of unification for these
values of $N_0$ are
\beq
\l N_0,~\fr{M_G}{\mu_0},~\fr{M_G}{\rm GeV},~\al_3 \r=
\l 1,~1.3 ,~1.7\cdot 10^{16},~0.117 \r~, ~~
\l 2,~2 ,~1.3\cdot 10^{16},~0.119 \r~.
\la{2unif321}
\eeq

The question may arise whether taking into account some threshold
corrections
will change the results or not. In fact, the SUSY threshold corrections
would introduce additional terms in (\ref{eq:A})-(\ref{eq:C}), which can
be important for the predictions of $\al_3$. These threshold corrections
can be
characterized by one 'threshold scale' $\hat{M}_{SUSY}$ and change
the strong coupling as \cite{nonunifMSSM1} 
${\al_3'}^{-1}=\al_3^{-1}+\fr{19}{28\pi }\ln 
\fr{\hat{M}_{SUSY}}{M_Z}$, where $\al_3^{-1}$ is given in (\ref{eq:A}).
For $100~{\rm GeV}<\hat{M}_{SUSY}<1$~TeV inequality (\ref{sazS}) is
modified to
$0.26\le S \le 3.13$, which would change $M_G$ not more than by factor 
of $3$. As we
see, there is no qualitative change, but scales can be modified
slightly. Because of this, throughout the paper we will not take into
account this type of threshold corrections.
Apart from this, on the 4D fixed points $N=1$ SUSY invariant kinetic
terms
are allowed, which in general could alter the unification picture
\cite{lockin}. However, if the size of extra dimension(s) is relatively
large
$1/R=\mu_0\ll M$($=$fundamental scale), then contributions from localized
kinetic terms will be negligible \cite{orbunif}. The condition $\mu_o\ll
M$
holds for the low scale unification scenarios considered below and is also
crucial for avoiding unwanted effects from other brane operators (see
Appendix A). 


%
%
%
%
%
%
%
%
%

{\bf Low scale unification}

By a look at the equations (\ref{eq:A}), (\ref{eq:B}) we recognize
that to have low 
scale unification it is important that the last term in eq.(\ref{eq:B}) is 
a large negative number, while the KK contributions in (\ref{eq:A}) must
be
small or vanish. Thus, to have low scale unification we need some
extension 
as to cancel the last term in (\ref{eq:A}) and to keep the
negative last term in (\ref{eq:B}). Among a few possible extensions
\cite{lowunif1}-\cite{lowunif3}, the simplest one seems to be the model of
ref. \cite{lowunif2}, where states 
${\bf E}_{N=2}^{(i)}= (E,~\ov{E})^{(i)}$ ($i=1, 2$) were introduced. 
The $E^{(i)}$,
$\ov{E}^{(i)}$ states are singlets of $SU(3)_c$, $SU(2)_L$ and carry
hypercharges $6$, $-6$, resp., in the $1/\sq{60}$ units
(\ref{hipcharge}). With the $Z_2$ parities 
\beq
(E^{(1)},~\ov{E}^{(2)})\sim +~,~~~~~
(E^{(2)},~\ov{E}^{(1)})\sim -~,
\la{Eparit}
\eeq
only the $E^{(1)}$, $\ov{E}^{(2)}$ states will have zero
modes. The contribution
of the ${\bf E}_{N=2}^{(1, 2)}$ states to the $\hat{b}$ factors  is then
\beq
\De_E (\hat{b}_1,~ \hat{b}_2,~\hat{b}_3)=\l \frac{12}{5},~0,~0\r~,
\la{BKKE}
\eeq
while the $b$ factors corresponding to the logarithmic runnings get
the additions
\beq
\De_E (b_1,~b_2,~b_3)=\l \fr{6}{5}, ~0,~0\r~ .
\la{BlogE}
\eeq
Taking these into account we have
%
\begin{equation}\label{eq:12}
                \alpha_3^{-1}= \frac{12}{7}\alpha_2^{-1}-\frac{5}{7}\alpha_1^{-1}+\frac{3}{7\pi}\ln{\frac{M_G}{M_E}},
\end{equation}
\begin{equation}\label{eq:13}
           \ln{\frac{M_G}{M_Z}}= \frac{5\pi}{14}(\alpha_1^{-1} - \alpha_2^{-1})-\frac{3}{14}\ln{\frac{M_G}{M_E}}-S,
\end{equation}
and
\begin{equation}\label{eq:14}
\alpha_G^{-1}=\alpha_2^{-1}-\frac{1}{2\pi}\ln{\frac{M_G}{M_Z}}+
\frac{1-2\eta}{\pi}S~,
\end{equation}
where $M_E$ is the 4D mass of $E^{(1)}$, $\ov{E}^{(2)}$ written as a
brane coupling.
In this setting the KK modes do not contribute in 
(\ref{eq:12}). To have a reasonable value for $\al_3$ one has to take
$M_E\simeq M_G$. The GUT scale $M_G$ can be as low as we like. With 
$M_G/\mu_0 \simeq 30$, $N_0=29$ we have $S=27.38$ and from
(\ref{eq:13}) we obtain
$M_G\simeq 25$~TeV. For $\eta =0$ in (\ref{eq:14}) the $\al_G$ remains in
the
perturbative regime. Although the value of $M_E$ is much  higher than
$\mu_0$, the form of the power law function $S$ of (\ref{DeKKZ}) is not
affected by the
$M_EE^{(1)}\ov{E}^{(2)}$ brane coupling. In appendix A we study the
possible brane operator effects on RGE and show that for 
$\mu_0\ll M_G$ they do not affect the expressions of
(\ref{DeKKZ}), (\ref{SKK}). Therefore the analysis carried out through
eqs.
(\ref{eq:12})-(\ref{eq:14}) remains valid.

\locsection{5D SUSY $SU(5)$ GUT on $S^{(1)}/Z_2\times Z_2'$ orbifold}

We start our study of GUT orbifold models with the 5D $N=1$ SUSY
$SU(5)$ theory. The fifth dimension is compact and  is considered to be
an
$S^{(1)}/Z_2\tm Z_2'$ orbifold. Two $Z_2$'s are necessary to avoid extra
zero mode states. In the notation of sect. 2, the 4D $N=1$ gauge
supermultiplet $V(24)$ in $G_{321}$ terms splits up as
\beq
V(24)=V_c(8,~1)_0+V_{SU(2)_L}(1,~3)_0+V_s(1,~1)_0+
V_X(3,~\bar 2)_{5}+V_Y(\bar 3,~2)_{-5}~,
\la{dec24ofSU5}
\eeq
where subscripts are the hypercharge $Y$ in the $1/\sq{60}$
units of (\ref{hipcharge}).
The decomposition of $\Si (24)$ will be similar to (\ref{dec24ofSU5}).

We ascribe to the fragments of $V(24)$ and $\Si (24)$
the following $Z_2\tm Z_2'$ parities
$$
\l V_c,~V_{SU(2)_L},~V_s\r \sim (+,~+)~,~~
\l V_X,~V_Y\r \sim (-,~+)~,
$$
\beq
\l \Si_c,~\Si_{SU(2)_L},~\Si_s\r \sim (-,~-)~,~~
\l \Si_X,~\Si_Y\r \sim (+,~-)~.
\la{Z2Z2ofV24}
\eeq
With this assortment all the couplings in
(\ref{lagv}) remain
invariant. From the $N=2$ SUSY $SU(5)$ gauge
supermultiplet only the $N=1$ gauge superfields of $G_{321}$ have zero
modes.
Therefore, at the $y=0$ fixed point (brane) we have a $N=1$ SUSY $G_{321}$
gauge theory.
The other states, including the $X$, $Y$ gauge bosons which would induce
$d=6$ nucleon decay, are projected out.
 
To have two MSSM Higgs doublets, one should introduce two $N=2$
supermultiplets ${\bf H}_{N=2}=(H,~\ov{H})$, 
${{\bf H}'}_{N=2}=(H', \ov{H}')$ where $H$, $H'$ are $5$-plets of $SU(5)$.
In terms of $G_{321}$ 
\beq
H(5)=h_u(1,~2)_{-3}+T (3,~1)_2~,~~
\ov{H}(\bar 5)=h_d (1,~\bar 2)_{3}+\ov {T}(\bar 3,~1)_{-2}~,
\la{higgsdec}
\eeq
and similarly for $H'$, $\ov{H}'$.
$\ov{H}$ and $\ov{H}'$ are mirrors of  $H$ and $H'$ resp.
With the following assignment of orbifold parities
\beq
(h_u,~{h_d}')\sim (+,~+)~,~~(h_d,~{h_u}')\sim (-,~-)~,~~
(T,~\ov{T}')\sim (-,~+)~,~~(\ov{T},~ T')\sim (+,~-)~,
\la{Z2Z2higgs}
\eeq
the states $h_u$, ${h_d}'$  have zero modes, which
we identify with one
pair of MSSM Higgs doublets. As we can see, all colored triplet states are
projected out and therefore will not participate in the $d=5$ nucleon
decays.
All couplings in (\ref{lagf}) are invariant under the $Z_2\tm Z_2'$
symmetry 
except the 
$M_{\Phi }\ov{\Phi }\Phi $ type couplings, which thus are not allowed.
This means that on the 5D level the $h_u{h_d}'$ coupling is absent due to
$N=2$ SUSY, while
$h_uh_d$ and ${h_u}'{h_d}'$ are not allowed by the orbifold parities.

Concerning the matter sector in $SU(5)$ we have anomaly free
$10$, $\bar 5$ multiplets, one per generation.
If at the level of the  5D SUSY theory we wish to introduce them as
bulk fields, we should
embed them into the $N=2$ matter supermultiplets. Per generation
we then have ${\bf {\cal X}}_{N=2}=(10,~ \ov{10})$ and 
${\bf \ov{\cal V}}_{N=2}=(\bar 5,~5)$, where $\ov{10}$ and $5$ are mirrors
of
$10$ and $\bar 5$ resp. In terms of $G_{321}$ this reads

\beq
10=e^c(1, 1)_{-6}+q (3, 2)_{-1}+u^c (\bar 3, 1)_{4}~,~~~
\bar 5=l (1, \bar 2)_3+d^c (1, \bar 3)_{-2}~.
\la{matterdec}
\eeq
Attempting to assign appropriate orbifold parities to the mirrors
and to project them out, one can easily realize that due to the parities
(\ref{Z2Z2ofV24}) 
of the gauge fields,  some of the states in (\ref{matterdec}) will not have zero
modes. To overcome this difficulty one can introduce copies 
(\cite{orbunif} and first refs. in \cite{orbunif1})
${{\bf {\cal X}}\hs{0.3mm}'}_{N=2}$, 
${{\bf \ov{\cal V}}\hs{0.3mm}'}_{N=2}$ where exactly 
those  states are allowed to have zero modes which correspond to the MSSM
states which come from ${\bf {\cal X}}_{N=2}$ and
${\bf \ov{\cal V}}_{N=2}$ and are projected out.
With orbifold parity prescriptions
\beq
\l q,~l,~{u^c}',~{d^c}', ~{e^c}' \r \sim (+,~+)~,~~
\l q',~l',~{u^c},~{d^c}, ~{e^c} \r \sim (-,~+)~,
\la{Z2Z2fer}
\eeq
and opposite ones for the corresponding mirrors, it is easy to verify that
now
the
scenario is compatible with the bulk construction since all terms (except
the 
mass term) of (\ref{lagf}) are invariant.

An alternative possibility would be to introduce fermionic states
only on the $y=0$ fixed point brane. This case can appear
in string models with intersecting branes \cite{branint}.
Thus, in general one can have $3-\eta $ of generations living at the
brane only and $\eta $ generations living also in the bulk. For the latter
case we have to
introduce $\eta $ copies. This applies not only for the $SU(5)$ model, but
also for the other scenarios considered below.
 
{\bf Fixed point brane couplings at $y=0$

and some phenomenology }

At the $y=0$ fixed point we are left with a SUSY $G_{321}$ gauge theory
with 
states $q$, $l$, ${u^c}'$, ${d^c}'$,
${e^c}'$, $h_u$, ${h_d}'$, which is  just the field content of
the MSSM. Since the 5D
action does not provide any Yukawa couplings, we should write appropriate
couplings on the brane. The
4D Yukawa superpotential, responsible for the generation of up-down quark
and lepton masses, has the form

\beq
W_Y=q{u^c}'h_u+q{d^c}'{h_d}'+l{e^c}'{h_d}'~.
\la{yukSU5}
\eeq 
Since states with primes and without primes come from different
unified
multiplets, in this case we do not have any asymptotic relations between
fermion masses. This avoids the problem {\bf (ii)} of fermion masses which
exists
in the
minimal $SU(5)$ GUT.
Also, since the colored triplets are projected out, the DT splitting
problem {\bf (iii)} as well as the problem {\bf (i)} caused by $d=5$
colored triplet exchange nucleon decay do not
exist any more.

However, the problem due to general $d=5$ baryon number violating
operators
and the $\mu $ problem are still unresolved on the 4D level 
(as in the case of the MSSM) unless some additional mechanism is applied.
One way for resolving these problems is to impose a continuous $R$
symmetry \cite{orbunif}, \cite{44model}, which in orbifold constructions
emerges on the
4D level after compactification as an $U(1)_R$ symmetry
\cite{orbunif}. The
latter can guarantee baryon number conservation, a suppressed $\mu $ term
and
automatic matter parity. Here and throughout the paper, alternatively, we
impose a discrete ${\cal Z}$ symmetry which allows some matter
parity/lepton
number violating operators responsible for the generation of appropriately
suppressed neutrino masses. In the spirit of sect. 4, we also
introduce  
${\cal S}$, $\ov{\cal S}$ singlets. With the transformations
(\ref{Ssym}) for 
$\ov{\cal S}{\cal S}$ 
and for doublets (\ref{mutermMSSM}) (here in all 
couplings $h_d$ must be replaced by ${h_d}'$) the relevant couplings will
be precisely the same as in 
(\ref{supmuMSSM}) and (\ref{supS}).
Through the couplings in (\ref{potS}) the VEV in
(\ref{solS}) is
obtained and consequently a $\mu $ term of the (\ref{valmuterm})
right magnitude is generated.

Since the minimal SUSY $SU(5)$ does not involve right handed neutrino
states
the neutrinos are massless. To give them mass, we also include,
together with the
Yukawa couplings (\ref{yukSU5}), the lepton
number violating bi-linear coupling 
(\ref{bilinS}), which (as discussed in sect. 4) induces the neutrino mass
(\ref{nubilinS}).
For $k=2$, $\lan {\cal S}/M\ran \sim 10^{-3}$ according to 
(\ref{bilinopS}) 
we obtain $m_{\nu_{\tau }}\sim 0.1$~eV, a
desirable value to explain the atmospheric anomaly.

With the couplings in (\ref{yukSU5}), (\ref{supmuMSSM}), 
(\ref{supS}), (\ref{bilinS}) and taking into account the unified $SU(5)$
multiplets on 5D level 
we have

$$
\al (\ov{\cal S})=\al-\al ({\cal S}) ~,~~~
\al ({h_d}')=\al -\al (h_u)~,
$$
$$
\al ({u^c}')=\al ({e^c}')=k\al ({\cal S})+2\al (h_u)-2\al ~,~~  
\al ({d^c}')=k\al ({\cal S})+4\al (h_u)-3\al ~,~~ 
$$
\beq
\al (l)=-k \al ({\cal S})-\al (h_u)+\al~,~~
\al (q)=-k \al ({\cal S})-3\al (h_u)+2\al ~,~~
\al =\fr{2\pi }{n}~,
\la{Zphases}
\eeq
where $\al (h_u) $, $\al ({\cal S})$ are undetermined. This still allows
the lepton number violating couplings (\ref{trilinS}) , giving
for $k=2$, $\lan {\cal S}\ran/M\sim 10^{-3}$ a radiative neutrino
mass $m_{\nu }' \sim 3\cdot 10^{-6}$~eV [see eq. (\ref{numastrilin}),
(\ref{trilinS1})].
This is the relevant scale for explaining the solar neutrino puzzle
through the large angle vacuum oscillations
of the $\nu_{e}$ state into the $\nu_{\mu, \tau}$.

As we will see in the next subsection, the scale of unification is close
to $10^{16}$~GeV. Assuming that the cutoff scale $M$ has the same
magnitude for
${\cal S}/M\sim 10^{-3}$ in (\ref{solS}) we need $n=3$. 
If in the eqs. (\ref{Zphases}) one takes
$k=2$, $n=3$, $\al (h_u)=\al/4 $, $\al ({\cal S})=\al/3$ 
the discrete symmetry would be $Z_{12n}$ and
one can verify that
baryon number violating $d=5$ operators $qqql$, 
${u^c}'{u^c}'{d^c}'{e^c}'$ are forbidden. Also, all other $R$-parity
and baryon number violating couplings are absent in this
scenario.

\subsection{Gauge coupling unification in 5D SUSY $SU(5)$}

Below the compactification scale $\mu_0$, we have precisely the MSSM
field content
with the $b$ factors given in (\ref{Bmssm}), while above $\mu_0$, due
to the
$Z_2\tm Z_2'$ parities of the states given in (\ref{Z2Z2ofV24}),
(\ref{Z2Z2higgs}) and for
$\eta $ generations of (\ref{Z2Z2fer}) in the bulk, we have

$$ 
\l \ga^{~}_1,~ 
\ga^{~}_2,~\ga_3\r =\l \fr{6}{5},~-2,~-6\r+
4\eta (1,~1,~1)~,
$$ 
\beq 
\l \de_1,~\de_2,~\de_3\r =
\l -\fr{46}{5},~-6,~-2 \r+
4\eta (1,~1,~1)~.
\la{BKKsu5} 
\eeq
{}From this and (\ref{als})-(\ref{alG}), and taking 
into account (\ref{DeKKZZ1}), we  get
\beq 
\al_3^{-1}=\fr{12}{7}\al_2^{-1}-\fr{5}{7}\al_1^{-1}-
\fr{6}{7\pi }(S_1-S_2)~,
\la{susysu5als} 
\eeq 
\beq 
\ln \fr{M_G}{M_Z}=\fr{5\pi }{14}(\al_1^{-1}-\al_2^{-1})- 
\fr{4}{7}(S_1-S_2)~,
\la{susysu5scale} 
\eeq 
\beq 
\al_G^{-1}=\al_2^{-1}-\fr{1}{2\pi }\ln \fr{M_G}{M_Z}+
\fr{1-2\eta}{\pi }S_1+\fr{3-2\eta}{\pi }S_2~.
\la{susysu5alG}  
\eeq
(Here we do not have an intermediate scale and we take $M_I=M_G$.) 
We see that contributions to (\ref{susysu5als}), (\ref{susysu5scale}) 
from the power law functions $S_1$, $S_2$ [defined
in (\ref{SKK})] are canceled out in the limit $S_1=S_2$. This is
understandable, since in this limit the $SU(5)$ symmetry is restored above
the $\mu_0$ scale and there are only contributions from complete $SU(5)$
multiplets
[according to (\ref{BKKsu5}) $\ga_a+\de_a=$const.]. To have a reasonable
value for $\al_3$ one needs $S_1-S_2\simeq 0$, which means
$M_G/\mu_0\simeq 1$ leading to $S_1\simeq S_2\simeq 0$. Because of this,
from (\ref{susysu5scale}) we get $M_G\simeq 2\cdot 10^{16}$~GeV. The value
of
$\al_G$ in (\ref{susysu5alG}) remains perturbative for $\eta=0-3$. Thus in
contrast to the  5D SUSY $G_{321}$ scenario, 
it is impossible to get low scale unification within the orbifold $SU(5)$
scenario.

\locsection{Step by step compactification and\\
power law unification}

In the previous section we have seen that within orbifold $SU(5)$ GUT a
power law unification does not take place. Although above  the $\mu_0=1/R$
scale each coupling of $G_{321}$ has power law running, the
renormalization of their relative slope (e.g. running of 
$\al_i^{-1}-\al_j^{-1}$) is still logarithmic because above  $\mu_0$ the
full $SU(5)$ multiplets participate in renormalization. Because of this,
one does not get low scale unification. This result
would be the same for
any higher dimensional orbifold GUT scenario with semisimple gauge group
[such as $SO(10)$, $SU(5+N)$, $E_6$, $E_8$, $\cdots $ ] if
the compactification of all extra dimensions occurs at a single mass
scale. Then
representations of all gauge
groups listed above again can be decomposed to complete $SU(5)$
multiplets.  Low scale unification 
within
GUT models with orbifold extra dimensions is however possible if we
allow for
compactifications of various extra dimensions at 
different mass scales. Suppose, we have a GUT model with gauge group $G$
and with
two extra spacial dimensions, and assume compactification in two steps
with scales 
$\fr{1}{R'}\gg \fr{1}{R}$ with a symmetry breaking chain

\beq
G\stackrel{1/R'}{\longrightarrow}  {\cal H}
\stackrel{1/R}{\longrightarrow}  {\cal H}_1~.
\la{chain}
\eeq
{}For having power law unification it is crucial  that  ${\cal H}$ must
be different from $SU(5)$ and also must not contain it as a subgroup.
Then the  field content of ${\cal H}$, relevant between
$\fr{1}{R'}$ and $\fr{1}{R}$, will not constitute full
$SU(5)$ multiplets. This would give us the possibility of low scale
unification. The bottom-up picture of such a scenario looks as follows: at
an energy scale $\mu_0=\fr{1}{R}$ the gauge group ${\cal H}_1$ is
'restored' to ${\cal H}$  and above $\mu_0$ KK states in incomplete
multiplets  give power law
differential running (of $\al_i^{-1}-\al_j^{-1}$). The gap 
$\fr{1}{R'}\gg \fr{1}{R}$ must be big enough to reduce the 'intermediate' 
scale
$\mu_0$ [see eq. (\ref{scale}), which gives a low intermediate scale $M_I$
in
case of large  $S$ (or $S_1$, $S_2$) and a negative coefficient; a
similar expression can be derived for scale $\mu_0$ in case of several
compactification mass scales]. Scale $\fr{1}{R'}$ is close to 
$M_G$: note, that the case with $\fr{1}{R'}\ll M_G$ will not work because
above $\fr{1}{R'}$ the unification group $G$ is restored and its full
multiplets would be 'alive'. 

In order to realize this idea, the gauge group $G$ {\it a)} must be higher
(in rank) than $SU(5)$; {\it b)} should have subgroups different from
$SU(5)$ and {\it c)} its subgroups must give a realistic phenomenology,
e.g. they should contain the $G_{321}$ gauge group and MSSM states. One of
the
groups, which has these properties, is $SO(10)$. Its maximal subgroups
are $SU(4)_c\tm SU(2)_L\tm SU(2)_R\equiv G_{422}$ and flipped 
$SU(5)\tm U(1)\equiv G_{51}$. It is straightforward to study
compactification breaking of these groups and to see what is going on
above 
the scale $\mu_0=\fr{1}{R}$. 
Since for power law unification the region between
$\fr{1}{R}$ and $\fr{1}{R'}\simeq M_G$ is relevant, we can consider five
dimensional $G_{422}$ and $G_{51}$ orbifold models. As we will see in the
following sections this type of bottom-up approach turns out to be quite
transparent and convenient for studying various phenomenological issues
together with gauge coupling unification.

\locsection{5D  $SU(4)_c\times SU(2)_L\times SU(2)_R$ $N=1$ SUSY model\\
on $S^{(1)}/Z_2\times Z_2'$ orbifold}


In the following we consider a supersymmetric 
$SU(4)_c\times SU(2)_L\times SU(2)_R\equiv G_{422}$ model
in five dimensions (5D).
$V(15_c) $ the adjoint of $SU(4)_c$,  in $SU(3)_c\times U(1)'$ terms reads
\beq
V(15_c)=V_s(1)_0+V_c(8)_0+V_t(3)_4+V_{\bar t}(\bar 3)_{-4}~,
\la{su4dec}
\eeq
where subscripts denote $U(1)'$ charges in $1/\sq{24}$ units ($SU(4)_c$
normalization):
\beq
Y_{U(1)'}=\fr{1}{\sq{24}}{\rm Diag}(1~,~ 1~,~ 1~,~ -3)~.
\la{u1charge}
\eeq
The decomposition of $\Si (15_c)$ is identical.

The decomposition of the $SU(2)_R$'s adjoints through the $SU(2)_R\to
U(1)_R$
channel has the form
\beq
V(3_R)=V_R(1)_0+V_p(1)_{2}+V_{\bar p}(1)_{-2}~,
\la{su2rdec}
\eeq
and the same for $\Si (3_R)$.
Here subscripts denote $U(1)_R$ charges in $1/2$ units:
\beq
Y_{U(1)_R}=\fr{1}{2}{\rm Diag}(1~,~ -1)~.
\la{u1rcharge}
\eeq

\paragraph{Matter sector.}We introduce $\eta $ generations of $N=2$ 
chiral supermultiplets 
\beq
{\bf F}_{N=2}=(F~, ~\ov{F})~,~~~
{\bf F^c}_{N=2}=(F^c~,~\ov{F}^c)~,
\la{matter}
\eeq
where under $G_{422}$
\beq
F\sim (4~,~2~,~1)~,~~~~F^c\sim (\bar 4~,~1~,~2)~,
\la{matterrep}
\eeq
\beq
F=(q~,~l)~,~~~~F^c=(u^c~,~d^c~,~\nu^c~,~e^c),
\la{content}
\eeq
and also  $\eta $ generations of  'copies' 
\beq 
{\bf F'}_{N=2}=(F'~, ~\ov{F}')~,~~~
{\bf F'^c}_{N=2}=(F'^c~,~\ov{F}'^c)~.
\la{mattercopy}
\eeq
with precisely the same content
and transformation properties as 
${\bf F}_{N=2}$ and ${\bf F^c}_{N=2}$ resp.
The remaining $3-\eta $ generations at the fixed point have the same
massless field content as the $\eta $ bulk generations.
Note that the introduction of copies is crucial if one wants a $Z_2\times Z_2'$ orbifold
invariant 5D action, with matter both in the bulk and on a fixed point, 
which reduces at low energies to the chiral content of the  MSSM, 
extended only by right handed neutrino states.

\paragraph{Scalar sector.} We need two sets of scalars. First we 
introduce hypermultiplets,
which will contain the two Higgs doublet superfields of the MSSM.
\beq
{\bf \Phi }_{N=2}=(\Phi ~,~\ov{\Phi })~,~~~~
{\bf \Phi' }_{N=2}=(\Phi' ~,~\ov{\Phi }')~,
\la{bidoublets}
\eeq
where under $G_{422}$
\beq
\Phi \sim (1~,~2~,~2)~,~~~\ov{\Phi } \sim (1~,~\bar 2~,~\bar 2)~.
\la{decdoubl}
\eeq
$\Phi $ and $\ov{\Phi }$ have the field content
\beq
\Phi=(h_u~,~h_d)~,~~~~\ov{\Phi }=(\ov h_u~,~\ov h_d)~.
\la{phicont}
\eeq
$h_u$ and $h_d$ have the same quantum numbers as the MSSM higgses 
responsible for the generation of up and down quark masses resp.
${\bf \Phi'}_{N=2}$ is a copy of ${\bf \Phi}_{N=2}$.
and its
introduction is crucial for having vectorlike Higgs
content, keeping the theory  anomaly free.

To break the $G_{422}$ symmetry we use orbifold compactification. Since
this breaking does not reduce the rank of the group we should introduce some additional scalars to provide for the needed rank breaking
via Higgs mechanism. The model
with minimal field content thus possesses also the following two $N=2$
supermultiplets
\beq
{\bf H}_{N=2}=(H^c~,~\ov H^c)~,~~~~
{\bf H'}_{N=2}=({H^c}\hsp{0.2mm}'~,~{\ov{H}^c}\hsp{0.1mm}')~,
\la{brscalars}
\eeq
where
\beq
H^c\sim (\bar 4~,~1~,~2)~,~~~
\ov{H}^c\sim (4~,~1~,\bar 2)~,
\la{decbrscal}
\eeq
\beq
H^c=(u^c_H~,~d^c_H~,~\nu^c_H~,~e^c_H)~,~~~ 
\ov{H}^c=(\ov{u}^c_H~,~\ov{d}^c_H~,~\ov{\nu }^c_H~,~\ov{e}^c_H)~,
\la{brscalcont}
\eeq
and ${\bf H'}_{N=2}$ is a copy of ${\bf H}_{N=2}$.

\subsection{$G_{422}\to G_{321}$ via 
$SU(3)\tm SU(2)_L\tm U(1)_R\tm U(1)'$ compactification breaking and 
related phenomenology}

In this subsection we will show how the $G_{422}$ symmetry can be broken down
to the standard model gauge group $G_{321}$. By a special selection of 
boundary conditions on the $S^{(1)}/Z_2\times Z_2'$ orbifold, at a first
stage
$G_{422}$ symmetry can be broken down to the
$SU(3)_c\times SU(2)_L\times U(1)'\times U(1)_R\equiv G_{3211}$ symmetry
(where $U(1)'$ and $U(1)_R$ come from $SU(4)_c$ and $SU(2)_R$
resp.). 
With the $Z_2\times Z_2'$ parities presented in
Table \ref{t:422GUT}, the
$G_{422}$ symmetry reduces to $G_{3211}$ 
and $N=2$ SUSY reduces to $N=1$ SUSY. On the fixed point $y=0$ we 
have only
states with $Z_2\times Z_2'$ parities $(+,~+)$. We therefore have three
generations of $q,~ l\hs{0.6mm}',~u^c,~{d^c}',~{\nu^c}',~e^c$, and 
two MSSM Higgs doublets $h_u$, ${h_d}'$.
In addition, there are extra 'scalar' supermultiplets 
$\nu^c_H$, ${\ov{\nu}^c_H}'$,
$d^c_H$, ${\ov{d}^c_H}'$. 

To break  $U(1)'\times U(1)_R$ down to the standard $U(1)_Y$ we use 
the states
$\nu^c_H$, ${\ov{\nu}^c_H}'$, which have zero $U(1)_Y$ hypercharge. By
developing non
zero VEVs along their scalar components they induce the desired breaking.
With $\lan \nu^c_H\ran $, $\lan {\ov{\nu}^c_H}'\ran $ we have unbroken 
\beq
Y=-\sq{\fr{2}{5}}Y_{U(1)'}+\sq{\fr{3}{5}}Y_{U(1)_R}~,
\la{suppos}
\eeq
where $Y$ is given in $(\ref{hipcharge})$ in the
'standard' $SU(5)$ normalization.
The superposition orthogonal to (\ref{suppos}) corresponds to the
hypercharge of the broken rank.

\begin{table} 
\caption{$U(1)'$, $U(1)_R$ charges and $Z_2\times Z_2'$
parities of various fragments in 5D SUSY $G_{422}$ scenario. 
All mirrors of 'matter' and 'scalar' states presented here have opposite
charges and parities.} 

\label{t:422GUT}
$$\begin{array}{|c|c|c|c|} 

\hline  N=1~{\rm superfield} &\sq{24}\cdot Y_{U(1)'} &
2\cdot Y_{U(1)_R} &
Z_2\times Z_2' \\  

\hline 
\hline

V_c,~V_{SU(2)_L},~V_s,~V_R  & 0 &0 & (+,~+) \\ 

\hline

\Si_c,~\Si_{SU(2)_L},~\Si_s,~\Si_R&0 &0 &(-,~-) \\

\hline

V_t~,~V_{\bar t}& 4~,~-4 &0 &(- ,~+ ) \\

\hline  

\Si_t~,~\Si_{\bar t}&4~,~-4 &0 &(+ ,~- ) \\ 

\hline 

V_p~,~V_{\bar p}&0 &2~,~-2 &(- ,~+ ) \\

\hline 

\Si_p~,~\Si_{\bar p}&0 &2~,~-2 &(+ ,~- ) \\

\hline  
\hline

q~~,~~q\hs{0.6mm}'&1 &0 &(+ ,~+ ),~(-,~+) \\

\hline

l~~,~~l\hs{0.6mm}'&-3 &0 &(- ,~+ ),~(+,~+) \\

\hline

u^c~~,~~{u^c}'&-1 &1 &(+ ,~+ ),~(-,~+) \\

\hline

d^c~~,~~{d^c}'&-1 &-1 &(- ,~+ ),~(+,~+) \\

\hline

\nu^c~~,~~{\nu^c}'&3 &1 &(- ,~+ ),~(+,~+) \\

\hline

e^c~~,~~{e^c}'&3 &-1 &(+ ,~+ ),~(-,~+) \\

\hline 
\hline

h_u~~,~~{h_d}'&0 &-1,~1 &(+ ,~+ ) \\

\hline 

{h_u}'~~,~~{h_d}&0 &-1,~1 &(- ,~+ ) \\

\hline 

\nu^c_H~~,~~{\ov{\nu }^c_H}'&3,~-3 &1,~-1 &(+ ,~+ ) \\

\hline

d^c_H~~,~~{\ov{d}^c_H}'&-1,~1 &-1,~1 &(+ ,~+ ) \\

\hline

e^c_H~~,~~{\ov{e}^c_H}'&3,~-3 &-1,~1 &(- ,~+ ) \\

\hline

u^c_H~~,~~{\ov{u}^c_H}'&-1,~1 &1,~-1 &(- ,~+ ) \\





\hline

\end{array}$$ 

\end{table}

Below we use a mechanism similar to the one we used in the MSSM and
$SU(5)$ cases, to generate the $\lan \nu^c_H\ran $ and 
$\lan {\ov{\nu }^c_H}'\ran $ VEVs. Here, to solve the various
phenomenological issues, we do not need to introduce the singlets ${\cal S}$, 
$\ov{\cal S}$ since their role will be played by   
$\nu^c_H$ and ${\ov{\nu }^c_H}'$. We introduce a ${\cal Z}$ symmetry 
under which the combination ${\ov{\nu }^c_H}'\nu^c_H$
($G_{3211}$ invariant) has
the same transformation as the $\ov{\cal S}{\cal S}$ in (\ref{Ssym}).
Then the relevant soft breaking terms and
consequently the whole potential is precisely the same as
in (\ref{supS}), (\ref{softSpot}), (\ref{potS}), but now with
${\cal S}$ and $\ov{\cal S}$ replaced by $\nu^c_H$
and ${\ov{\nu }^c_H}'$.
Also, as in (\ref{solS}), the solutions $\lan \nu^c_H\ran$, 
$\lan {\ov{\nu }^c_H}'\ran $ with parameterization
\beq
\lan \nu^c_H\ran \sim \lan {\ov{\nu }^c_H}'\ran\equiv v~,
\la{nuvevdef}
\eeq
will be
\beq
v\sim
M\l \fr{m}{M}\r^{\fr{1}{2n-2}}~.
\la{nusol}
\eeq
In this way the $U(1)'\times U(1)_R$ symmetry breaking scale $v$
is expressed by the interplay of the cutoff scale $M$ and the SUSY mass
scale
$m$, and the magnitude of $v$
is controlled by the discrete ${\cal Z}$ symmetry, i.e. by $n$. 
Depending on the scenario one considers, one can select $n$ in such a way as to obtain
a reasonable ratio $v/M$. For instance for $n=4$, $m\simeq 1$~TeV, 
$M\simeq M_G\sim 10^{13}$~GeV we get $v/M\sim 2\cdot 10^{-2}$, which is
indeed a desirable value (see first row of Table 
\ref{t:Iunif422}
which
corresponds to model {\bf I}-susy422). 
{}For low values of $M$, there are no large mass gaps, and 
there is no need for large $n$'s: e.g. for  $m=500$~GeV, 
$M\sim 500$~TeV, and $n=2$
we have $v/M\sim 0.01$. Also this case can be realized with successful
unification of gauge couplings (see rows 1-3 of Table \ref{t:IIIunif422},
corresponding to the model {\bf III}-susy422).


{}To avoid the 4D superpotential coupling $Mh_u{h_d}'$ we postulate 
the transformation property (\ref{mutermMSSM}), where $h_d$ is to be
replaced by
${h_d}'$.
The coupling responsible for the $\mu $ term generation then is
\beq
W_{\mu }=M\l \fr{{\ov{\nu }^c_H}'\nu^c_H}{M^2}\r^{n-1}h_uh_d'~,
\la{musup}
\eeq
and after substituting the VEVs of the $\nu^c_H$, ${\ov{\nu }^c_H}'$ states,
taking  into account (\ref{nusol}), we get
\beq
\mu\sim M\l \fr{v}{M}\r^{2n-2}\sim m~.
\la{muterm}
\eeq

%

For the time being the introduced discrete symmetry is acting  
as $Z_n$ on the field combinations ${\ov{\nu }^c_H}'\nu^c_H$, $h_u{h_d}'$ 
while the transformation properties of the single fields were not specified. 
The phases of the ${\cal Z}$ transformations for the introduced states
are given  in Table \ref{t:Z422}.  The 4D Yukawa
superpotential generating charged fermion masses reads
%
%
\begin{table} \caption{$\phi_i\to e^{{\rm i}\al (\phi_i)}\phi_i$
transformation
properties of various superfields. $\om =\fr{2\pi }{12n}$, while 
$\al (q)$ 
is a free phase and we take $\al (q)=5\om $.}

\label{t:Z422} $$\begin{array}{|c|c|c|c|c|}
 
\hline 
{\rm Field}~\phi_i &~{\nu^c}_H~&~{\ov{\nu^c }_H}'~  &~h_u~&~{h_d}'~   \\
 
\hline 
 
\al (\phi_i) &-\al (q)+3(4-k)\om  & \al (q)+3k\om   &3(2-k)\om  &
3(2+k)\om \\
 
\hline \hline

{\rm Field}~\phi_i &~q~ &~l'~ &u^c,~e^c &{d^c}',~{\nu^c}'  \\

\hline

\al (\phi_i) &\al (q) & \al (q)-6k\om &
-\al (q)-3(2-k)\om &-\al (q)-3(2+k)\om  \\
\hline

\end{array}$$
 
\end{table} 
%
%
\beq
W_Y=qu^ch_u+q{d^c}'{h_d}'+e^cl'{h_d}'~,
\la{yuksup}
\eeq
where family indices are suppressed.

{}For the Dirac and Majorana couplings of neutrinos  we
have
\beq
W_{\nu }=\lam_{\nu }\l \fr{{\ov{\nu }^c_H}'{\nu^c}_H}{M^2}\r^k
{\nu^c}'l'h_u+
\fr{{\ov{\nu }^c_H}'{\nu^c}_H}{M^3}({\ov{\nu }^c_H}'{\nu^c}')^2~,
\la{dirmaj422}
\eeq
where $k$ is some integer, the value of which is dictated by 
the model considered, 
depending on what suppressions for neutrino masses are needed.
After substituting the appropriate VEVs and integrating out the ${\nu^c}'$
states
(with masses $M_{{\nu^c}'}\sim v^4/M^3$), we obtain neutrino masses 
\beq
m_{\nu }\simeq \lam_{\nu }^2\fr{h_u^2}{M}
\l \fr{v}{M} \r^{4k-4}~.
\la{numass422}
\eeq
{}For $\lam_{\nu }\sim 1$, $k=1$, $M\sim 10^{13}$~GeV this gives 
$m_{\nu }\sim h_u^2/M\sim 1$~eV.
In the case $k=2$, $m_{\nu }\sim \l\fr{v}{M}\r^4\fr{h_u^2}{M}$
it is possible to reduce the cut off scale even down to the multi
TeV: for $M\sim 100$~TeV and $v/M\sim 10^{-2}$ we get 
$m_{\nu }\sim 1$~eV.
We will see below that both cases $k=1,~2$ can be
realized and
give successful pictures of unification
(see models {\bf I}-susy422 and {\bf III}-susy422 in
sect. 7.1.1). 
In order to accommodate recent atmospheric data with neutrino mass scales
$1$~eV, one might assume neutrino species with degenerate masses
\cite{degnu}, e.g. $\sum_i m_{\nu_i}\sim 3$~eV so that neutrinos are
candidates for dark matter. 
An alternative solution of the atmospheric and solar neutrino
puzzles could be provided by a hierarchical structure of masses, which
requires
$m_{\nu_3}\sim 0.1$~eV. This scale can be obtained having
$\lam_{\nu_3}\sim 1/3$ in (\ref{numass422}). A stronger suppression for
the first
two neutrino generations can be achieved by introducing some 
flavor symmetries in the spirit of \cite{nufl}.

Matter $R$-parity violating operators
\beq
{\nu^c}_H h_u l'~,~~~{\nu^c}_H q {d^c}' l' ~,~~~
{\nu^c}_H u^c {d^c}' {d^c}'~,~~~
{\nu^c}_H e^c l' l'~,
\la{parviol}
\eeq
which are invariant under $G_{3211}$, are forbidden by the $Z_{12n}$ symmetry
with the charge selections given in Table \ref{t:Z422}, for $k=1$ or
$2$. The $d=5$ baryon number violating operators of the type presented 
in (\ref{d5MSSM}) and also $qqq{h_d}'{\ov{\nu^c}_H}'$ 
(this operator violates also $R$ parity and leads to a $d=5$ baryon number
violating coupling after
the substitution of VEV $\lan {\ov{\nu }^c_H}'\ran $)
are forbidden in this scenario for both choices
$k=1$ and $k=2$.
There are also $d=6$ (\ref{d6MSSM}) type operators 
allowed by $G_{3211}$ symmetry and in addition the
$(qq{{d^c}'}^{\hs{0.05cm}+}{{\nu^c}'}^{\hs{0.05cm}+})_D$ coupling.
It is easy to check that these are absent due to the $Z_{12n}$ symmetry.

In order that the 5D Lagrangian terms of (\ref{lagv}), (\ref{lagf}),
allowed by
the $Z_2\tm Z_2'$ orbifold parities, are invariant under the
introduced $Z_{12n}$ discrete symmetry, we must assure that the other fields transform properly.
This is the case if
$\al (F)=\al (q)$, $\al (F')=\al (l')$,
$\al (F^c)=\al (u^c)$, 
$\al({F^c}')=\al ({d^c}')$,
$\al (\Phi )=\al (h_u)$, 
$\al ({\Phi }')=\al ({h_d}')$,
$\al (H^c)=\al (\nu^c_H)$, 
$\al ({\ov{H}^c}')=\al ({\ov{\nu}^c_H}')$
(and all mirrors with opposite phases).

Since the states 
$d^c_H$ and ${\ov{d}^c_H}'$ transform as $\nu^c_H$
and ${\ov{\nu }^c_H}$ resp., their mass term is
generated through an operator 
$({\ov{\nu }^c_H}'\nu^c_H/M^2)^{n-1}M{\ov{d}^c_H}'d^c_H$ and one gets 
$M_{d^c_H}\sim m$. First of all we must make sure that these triplet
states do not cause nucleon decay. The allowed couplings of 
$d^c_H$, ${\ov{d}^c_H}'$ with matter are
$({\ov{\nu }^c_H}'\nu^c_H)^{n-1}\nu^c_Hql'd^c_H$ and
${\ov{\nu }^c_H}'e^cu^c{\ov{d}^c_H}'$ for $k=1$, while
for $k=2$ the operators 
$\nu^c_Hql'd^c_H$ and 
$({\ov{\nu }^c_H}'\nu^c_H)^{n-1}{\ov{\nu }^c_H}'e^cu^c{\ov{d}^c_H}'$
are permitted. 
However, the couplings 
$\nu^c_Hu^c{d^c}'d^c_H$ and ${\ov{\nu }^c_H}'qq{\ov{d}^c_H}'$
are forbidden (see Table \ref{t:Z422}) and the baryon number violating 
$d=5$ operators $qqql$, $u^cu^c{d^c}'e^c$ do not emerge. 
The issue of gauge coupling unification in this model , which we call {\bf I}-susy422, will be studied below. As it turns out, successful
unification can be obtained for various scales as presented in 
Table \ref{t:Iunif422}. For the case shown in the first row we obtain  
$v/M\sim 2\cdot 10^{-2}$ (obtained for $n=4$ according to
(\ref{nusol})). This mass gap is crucial for $\mu $-term generation with
the
correct magnitude. For this case we have $M_{d^c_H}\simeq 10$~TeV. 
The existence of colored triplet states with this mass can have interesting
phenomenological implications \cite{lightTrip}. There might be a 
leptoquark
like signature \cite{leptq}, similar to what is expected within some 
$R$-parity violating models.

A different scenario, with heavy $d^c_H$, ${\ov{d}^c_H}'$ states, can be
constructed introducing additional two $N=2$ supermultiplets
$6^{(i)}_{N=2}=(6,~\bar 6)^{(i)}$ ($i=1, 2$) of $SU(4)_c$, where
\beq
6^{(i)}=T^{\hsp{0.2mm}(i)}(3_{-2})+\ov{T}^{\hsp{0.2mm}(i)}(\bar 3_2)~,~~~~
\bar 6^{(i)}={\cal T}^{\hsp{0.2mm}(i)}(3_{-2})+
\ov{\cal T}^{\hsp{0.3mm}(i)}(\bar 3_2)~.
\la{6pl}
\eeq 
With the $Z_2\times Z_2'$ parity assignment
$$
(T^{\hsp{0.2mm}(1)},~\ov{T}^{\hsp{0.3mm}(2)})\sim (+,~+),~~~~~
(T^{\hsp{0.2mm}(2)},~\ov{T}^{\hsp{0.2mm}(1)})\sim (-,~+)~,
$$
\beq
({\cal T}^{\hsp{0.2mm}(1)},~\ov{\cal T}^{\hsp{0.3mm}(2)})\sim
(+,~-)~,~~~~~
({\cal T}^{\hsp{0.2mm}(2)},~\ov{\cal T}^{\hsp{0.3mm}(1)})\sim (-,~-)~,
\la{trippar}
\eeq
the triplet-antitriplet pair $T^{\hsp{0.2mm}(1)}$, 
$\ov{T}^{\hsp{0.3mm}(2)}$ will have zero
modes and can therefore couple
with the $d^c_H$, ${\ov{d}^c_H}'$ giving them large masses.
With $Z_{12n}$ phases
$\al (T^{\hsp{0.2mm}(1)})=-2\al (\nu^c_H)$,
$\al (\ov{T}^{\hsp{0.3mm}(2)})= -2\al ({\ov{\nu}^c_H}')$
the relevant 4D superpotential couplings will thus be
\beq
W_T=\nu^c_Hd^c_HT^{\hsp{0.2mm}(1)}+
{\ov{\nu }^c_H}'{\ov{d}^c_H}'\ov{T}^{\hsp{0.3mm}(2)}~.
\la{tripsup}
\eeq
After substituting the VEVs of $\nu^c_H$, ${\ov{\nu }^c_H}'$, the triplet
states acquire masses $M_T\sim v$. The allowed couplings of the
$T^{\hsp{0.2mm}(1)}$, $\ov{T}^{\hsp{0.3mm}(2)}$
states with matter, are
$({\ov{\nu }^c_H}'\nu^c_H)^{k}ql'\ov{T}^{\hsp{0.3mm}(2)}$,
$({\ov{\nu }^c_H}'\nu^c_H)^{3-k}e^cu^cT^{\hsp{0.2mm}(1)}$.
However, the couplings 
$qqT^{\hsp{0.2mm}(1)}$, $u^c{d^c}'\ov{T}^{\hsp{0.3mm}(2)}$
are forbidden by $Z_{12n}$ symmetry and 
baryon number is still conserved.
We refer to this model as {\bf II}-susy422. Also in this case 
successful unification of gauge couplings occurs if $M_I\simeq \mu_0\simeq
M_G$
(see sect. 7.1.1). However, as we will see, with a specific extension 
it is possible to get unification near the multi TeV region
(see 
Table \ref{t:IIIunif422} for the model {\bf III}-susy422, which presents mass
scales for which unification holds).
For this case, since triplets get masses $\sim M_I$ through the
couplings (\ref{tripsup}), their masses are a few TeV, making this scenario
testable in future collider experiments.

We conclude this section by noting that, together with a natural
$U(1)'\times
U(1)_R$ breaking pattern and
$\mu $-term generation, the $Z_{12n}$ symmetry provides automatic
$R$-parity
and baryon number conservation within the 5D SUSY orbifold $G_{422}$
model.

\subsubsection{Gauge coupling unification in 5D SUSY 
with $G_{422}\to G_{3211}$\\
intermediate breaking}

Here we will study the issue of gauge coupling unification
for SUSY $G_{422}$ model with compactification breaking to
$G_{3211}$. Throughout this analysis we will 
use the expressions obtained in section 3.

{\bf Model I-susy422}

The field content of this scenario is as follows. We have the scalar
superfields of (\ref{bidoublets}), (\ref{brscalars}), which are necessary
to obtain the
pair of MSSM Higgs doublets and to realize the wanted 
$G_{3211}$ breaking  to $G_{321}$. 
We also have $\eta $ generations of  $F$, $F^c$
presented in
(\ref{matter}), and $\eta $ copies, if $\eta $ generations of matter 
have KK excitations.  We then
identify the scale of $U(1)_R\times U(1)'$ symmetry breaking 
$\lan \nu^c_H\ran =\lan {\ov{\nu}^c_H}'\ran $ with the intermediate scale 
$M_I$ in (\ref{RGE321}), (\ref{match}). Below $M_I$, the gauge group is
$G_{321}$ and the field content is that of the
MSSM with the $b$-factors (\ref{Bmssm}),
plus the states $d^c_H$, ${\ov{d}^c_H}'$ with a mass $M_{d^c_H}$ in the
range 
$100$~GeV-$1$~TeV, which have $b$-factors
\beq
b_i^{d^c_H}=\l \fr{2}{5}~,~0~,~1 \r~.
\la{Bdc}
\eeq
Above the scale $M_I$ we have 
\beq
\l b_{U(1)_R},~b_{U(1)'},~b_2,~b_3\r^{M_I} =\l 9,~7,~1,~-2\r~.
\la{B3211}
\eeq
With the $Z_2\times Z_2'$ parities shown in Table 
\ref{t:422GUT}, the corresponding $\ga $ and $\de $-factors of
(\ref{DeKKZZ1}),
will be  
$$
\l \ga^{~}_{U(1)_R},~\ga^{~}_{U(1)'},~
\ga^{~}_2,~\ga_3\r =\l 6,~2,~-2,~-4\r +
4\eta \l 1,~1,~1,~1 \r~,
$$
\beq
\l \de_{U(1)_R},~\de_{U(1)'},~\de_2,~\de_3\r =
\l 2,~-6,~2,~0 \r+4\eta \l 1,~1,~1,~1 \r~.
\la{BKK}
\eeq

Due to (\ref{suppos}), which determines the pattern of $U(1)_Y$ embedding
in
$U(1)_R\times U(1)'$, the group-theoretical factor $\tan \te $ in
(\ref{match}), (\ref{BsupG1G2}), (\ref{KKsupG1G2}) will be $\tan \te
=\sq{\fr{3}{2}}$ if
$G_1=U(1)_R$ and
$G_2=U(1)'$. Taken this into account, using (\ref{match})-(\ref{Delog}), 
(\ref{DeKKZZ1}), (\ref{KKsupG1G2}), it is not
difficult to derive from (\ref{als})-(\ref{alG}) the following equations 
\beq
\al_3^{-1}=\fr{12}{7}\al_2^{-1}-\fr{5}{7}\al_1^{-1}+
\fr{9}{14\pi }\ln \fr{M_I}{M_{d^c_H}}+ 
\fr{15}{14\pi }\ln \fr{M_G}{M_I}+
\fr{9}{7\pi }S_1-
\fr{15}{7\pi}S_2~,
\la{modelals}
\eeq
\beq
\ln \fr{M_I}{M_Z}=\fr{5\pi }{14}(\al_1^{-1}-\al_2^{-1})-
\fr{1}{14}\ln \fr{M_I}{M_{d^c_H}}- 
\fr{9}{7}\ln \fr{M_G}{M_I}-
\fr{8}{7}S_1+\fr{4}{7}S_2~,
\la{modelscale}
\eeq
\beq
\al_G^{-1}=\al_2^{-1}-\fr{1}{2\pi }\ln \fr{M_G}{M_Z}+
\fr{1-2\eta }{\pi }S_1-\fr{1+2\eta }{\pi }S_2~.
\la{modelalG}
\eeq
Without the four last terms in (\ref{modelals}) the one loop value of $\al_3$
would be $0.116$, which is close to the experimental value of the 
strong coupling $\al_3(M_Z)$. 
Therefore, the contribution of the remaining terms should not
be 
large. Since $S_1$ and $S_2$ have nearly the same values, 
the sum of the last two
terms in (\ref{modelals}) will be negative and this negative number must
be compensated by the third and fourth term by a proper choice of the 
mass scales. In (\ref{modelscale}) 
the last four terms give a negative contribution. This 
gives the possibility to have a relatively low scale $M_I$. This is
realized
for $\eta=0$. The latter is crucial for gauge constant's perturbativity.
Within all models with large gap between $\mu_0$ and $M_G$
scales we will take $\eta =0$, i.e. matter is located at the fixed point
(in agreement with string models with intersecting branes \cite{branint}).
Mass scales, which give successful
unification, are presented in Table \ref{t:Iunif422}.
$N$, $N'$ are the maximal numbers of even and odd KK states resp., 
which lie below $M_G$ 
and are determined from the inequalities (\ref{maxNs}).
The picture of unification for the $M_G/\mu_0=30$ case of Table
\ref{t:Iunif422} is presented in Fig. 1.

%
%

\begin{table} \caption{Unification for SUSY $G_{422}$ model
{\bf I}-susy422, with $\al_3(M_Z)=0.119$ and $\eta=0$. Mass scales are
measured in GeV units.}
 
\label{t:Iunif422} $$\begin{array}{|c|c|c|c|c|c|}
 
\hline 
M_G/\mu_0&(N ,~N') & \log_{10}[M_I] & 
\log_{10}[\mu_0] & \log_{10}[M_G]& \log_{10}[M_I/M_{d^c_H}]  \\
 
\hline \hline
 
30 &(14 ,~14 )  &10.93  &11.15   &12.63  &6.92  \\
 
\hline

28 & (13 ,~13) &11.5 &11.5  &12.94 &6.75 \\

\hline 

26&(12 , ~13) &11.64 &11.64 &13.08 &6.5 \\ 

\hline 

\end{array}$$
 
\end{table}

%
%

{\bf Model II-susy422}

In this scenario we introduce two $6_{N=2}$ supermultiplets with
the components shown in
(\ref{6pl}) and with $Z_2\times Z_2'$ parities shown in
(\ref{trippar}). As we will see, with only this extension, it is impossible
to get unification with a mass gap between $M_I$, $\mu_0$ and $M_G$, 
as well as relatively low $M_G$. But this can be achieved with a simple
additional extension (model {\bf III}-susy422).

Due to the
couplings (\ref{tripsup}), all triplet zero mode states decouple on the
$M_I=v$
scale and consequently, below $M_I$, the gauge coupling runnings will be
precisely the same as in the MSSM, with $b$-factors (\ref{Bmssm}). Due to 
the presence of the $T^{\hsp{0.2mm}(1)}$, $\ov{T}^{\hsp{0.3mm}(2)}$
states, the $b$-factors will be modified above $M_I$ and (\ref{B3211}) will
be changed by
\beq
\De_{\bf 6} \l b_{U(1)_R},~b_{U(1)'},~b_2,~b_3\r^{M_I} =
\l 0,~1,~0,~1\r~.
\la{6BIIsusy422}
\eeq
The $\ga $ and $\de $ factors in (\ref{BKK})  
are modified by
\beq
\De_{\bf 6} \l \ga^{~}_{U(1)_R},~\ga^{~}_{U(1)'},~
\ga^{~}_2,\ga_3\r =
\De_{\bf 6} \l \de_{U(1)_R},~\de_{U(1)'},~\de_2,~\de_3\r
=\l 0,~2,~0,~2\r~.
\la{6gdeIIsusy422}
\eeq
(The subscript '{\bf 6}' in $\De_{\bf 6}$ indicates that the changes
are due to states coming from the two ${\bf 6}_{N=2}$
supermultiplets.) Taking all this into account, we  have
\beq 
\al_3^{-1}=\fr{12}{7}\al_2^{-1}-\fr{5}{7}\al_1^{-1}+
\fr{12}{7\pi }\ln \fr{M_G}{M_I}+\fr{18}{7\pi }S_1-\fr{6}{7\pi }S_2~,
\la{IIsusy422als} 
\eeq 
\beq 
\ln \fr{M_I}{M_Z}=\fr{5\pi }{14}(\al_1^{-1}-\al_2^{-1})- 
\fr{19}{14}\ln \fr{M_G}{M_I}-\fr{9}{7}S_1+\fr{3}{7}S_2~.
\la{IIsusy422scale} 
\eeq 
{}In (\ref{IIsusy422als}) we see that the contribution of the last three
terms is always positive and for reasonable $\al_3$ the only possibility is to
have $M_I\simeq \mu_0 \simeq M_G$. 
Due to this fact, from (\ref{IIsusy422scale}) one
can see that $M_I\simeq 2\cdot 10^{16}$~GeV.

{\bf Model III-susy422: low scale unification}

Here we present an extension which gives low scale unification. 
In addition to the field content of model {\bf II}-susy422 we introduce
two ${\bf 2}_{N=2}^{(j)}=(D,~\ov{D})^{(j)}$ ($j=1, 2$), which are supermultiplets of $SU(2)_L$. 
With $Z_2\times Z_2'$ parities
\beq
(D^{\hsp{0.2mm}(1)},~{\ov{D}^{\hsp{0.4mm}(2)}})\sim (+,~+)~,~~~~~~
(D^{\hsp{0.2mm}(2)},~{\ov{D}^{\hsp{0.4mm}(1)}})\sim (-,~-)~,
\la{2par}
\eeq
only the states $D^{\hsp{0.2mm}(1)}$, ${\ov{D}^{\hsp{0.4mm}(2)}}$ 
have zero modes. The contributions to the $b^{M_I}$,
$\ga $, and $\de $-factors, due to ${\bf 2}_{N=2}^{(j)}$ supermultiplets,
are
$$
\De_{\bf 2}\l b_{U(1)_R},~b_{U(1)'},~b_2,~b_3\r^{M_I}=(0,~0,~1,~0)~,
$$
\beq
\De_{\bf 2}\l \ga_{U(1)_R},~\ga_{U(1)'},~\ga_2,~\ga_3\r 
=(0,~0,~2,~0)~,~~~~~~
\De_{\bf 2}\de_i=0~, 
\la{2BgadeIIsusy422}
\eeq
and consequently we obtain
\beq
\al_3^{-1}=\fr{12}{7}\al_2^{-1}-\fr{5}{7}\al_1^{-1}+ 
\fr{12}{7\pi }\ln \fr{M_G}{M_I}-
\fr{6}{7\pi }\ln \fr{M_G}{M_D}
+\fr{6}{7\pi }(S_1-S_2)~,
\la{lowIIsusy422als}
\eeq
\beq
\ln \fr{M_I}{M_Z}=\fr{5\pi }{14}(\al_1^{-1}-\al_2^{-1})-
\fr{19}{14}\ln \fr{M_G}{M_I}+
\fr{5}{28}\ln \fr{M_G}{M_D}
-\fr{13}{14}S_1+\fr{3}{7}S_2~,
\la{lowIIsusy422scale}
\eeq
\beq
\al_G^{-1}=\al_2^{-1}-\fr{1}{2\pi }\ln \fr{M_G}{M_Z}-
\fr{1}{2\pi }\ln \fr{M_G}{M_D}
-\fr{2\eta }{\pi }S_1-\fr{1+2\eta }{\pi }S_2~,
\la{lowIIsusy422alG}
\eeq
where $M_D$ is the mass of the zero mode of the doublet states 
$D^{\hsp{0.2mm}(1)}$, $\ov{D}^{\hsp{0.3mm}(2)}$, which arises from the 4D
superpotential coupling
$M_DD^{\hsp{0.2mm}(1)}\ov{D}^{\hsp{0.3mm}(2)}$.
{}From (\ref{lowIIsusy422als}) we see that for $S_1=S_2$ the contribution
from
KK states cancels out. Since $S_1$, $S_2$ differ slightly, the
cancellation is partial and for a desirable value of $\al_3(M_Z)$,
appropriate contributions from the logarithmic terms are needed. 
Now, the contribution from KK states in
(\ref{lowIIsusy422scale}) is
always negative (there is no possible cancellation). Thus it is
possible to get low scale unification in this scenario. The values of the mass
scales, which give a successful picture of unification, are presented
in Table \ref{t:IIIunif422}. As one can see, for 
$M_G/\mu_0=110\hs{-0.08cm}-\hs{-0.08cm}107$ the
couplings
unify at the scale 
$M_G\simeq (200\hs{-0.08cm}-\hs{-0.08cm}400)$\hs{0.07cm}TeV;  
the unified coupling constant 
$\al_G$($\simeq 0.1$) is perturbative. 
The picture of unification for case $M_G/\mu_0=110$ of Table
\ref{t:IIIunif422} is presented in Fig. 2.
Note, that in
these cases the mass of the doublet pair 
$D^{\hsp{0.2mm}(1)}$, $\ov{D}^{\hsp{0.3mm}(2)}$ 
is $M_D\simeq (123\hs{-0.08cm}-\hs{-0.08cm}257)$\hs{0.07cm}GeV. 
This makes the model testable in future collider experiments.

%
%

\begin{table} \caption{Unification for the SUSY $G_{422}$ model
{\bf III}-susy422, with $\eta=0$. In all cases $\al_G\simeq 0.1$ is
perturbative. 
Scales are measured in GeV units.}
 
\label{t:IIIunif422} $$\begin{array}{|c|c|c|c|c|c|c|}

\hline 
M_G/\mu_0&(N ,~N') & \log_{10}[M_I] & 
\log_{10}[\mu_0] & \log_{10}[M_G]&
\log_{10}[M_G/M_D]&\al_3(M_Z)    \\
 
\hline \hline
 
 110& (54  , ~54  ) &3.26  &3.26 &5.30 &3.21 &0.1184    \\
 
\hline

 108& (53 , ~53 ) &3.48  &3.48  &5.52  &3.21  &0.1185    \\

\hline

 107& (52  , ~53 ) &3.6 & 3.6  &5.62 &3.21 &0.1186   \\

\hline

 106& (52 , ~52 ) &3.71 &3.71 &5.73 &3.21 &0.1186 \\

\hline

\end{array}$$
 
\end{table}

%
%

\subsection{$G_{422}\to G_{321}$ via 
$SU(3)\hs{-0.1cm}\tm SU(2)_L\hs{-0.05cm}\tm SU(2)_R\hs{-0.05cm}\tm U(1)'$
compactification\\
breaking and related phenomenology}

In this subsection, similar to 7.1, we will consider the breaking of
$G_{422}$ by orbifold
compactification down to 
$SU(3)\tm SU(2)_L\tm SU(2)_R\tm U(1)'\equiv G_{3221}$. 
The breaking  of the latter to $G_{321}$ again will occur on the 4D level 
through the non-vanishing VEVs
of certain fields.  

The decomposition of $V(15_c)$ ($\Si (15_c)$) under $SU(3)_c\tm
U(1)'$ is given in (\ref{su4dec}). The gauge group $SU(2)_L\tm SU(2)_R$
is,
in this case, not broken by the compactification.

The matter sector is the same as in (\ref{matter}) with copies 
(\ref{mattercopy}) needed if chiral states are introduced in the bulk. The
content of $F$, $F^c$ from the viewpoint of $G_{3221}$ is
\begin{equation}
F=q(3,~2,~1)_{1}+l(1,~2,~1)_{-3}~,~~~ 
F^c = q^c(\bar 3 ,~1,~2)_{-1}+l^c(1,~ 1,~2)_{3}~,
\end{equation}
where
\begin{equation}
q^c=(u^c,~d^c),\quad l^c=(e^c,~\nu^c)~.
\end{equation}
For the copies we have similar expressions.

To obtain one pair of MSSM Higgs doublets, it is enough in this case to
have
only one $N=2$ supermultiplet ${\bf \Phi }_{N=2}=(\Phi ,~\ov{\Phi })$ 
(see (\ref{bidoublets})) with field content as in (\ref{decdoubl}),
(\ref{phicont}).

For further $SU(2)_R\times  U(1)'$ breaking to $U(1)_Y$ 
we need the states  (\ref{brscalars}) with components 
(\ref{decbrscal}), (\ref{brscalcont}) where now under
$G_{3221}$ $H^c$ decomposes as
\beq
H^c=q^c_H(\bar 3 ,~1,~2)_{-1}+l^c_H(1,~ 1,~2)_{3}~
\la{dec3221}
\eeq
and similarly for $H'$, $\ov{H}^c$, ${\ov{H}^c}'$.
The $Z_2\tm Z_2'$ parities and $U(1)'$ charges of the appropriate fragments
are given in Table \ref{t:LR3}. With these parity assignments, $G_{422}$
is broken to $G_{3221}$ and the zero mode matter and 'scalar' superfields
are $q$, $l'$, $q^c$, ${l^c}'$, $\Phi $, $l^c_H$, 
${\ov{l}^c_H}'$.

\begin{table} \caption{$U(1)'$ charges and $Z_2 \times Z_2'$ parities of the various fragments. All mirrors have opposite charges and parities.}

\label{t:LR3} $$\begin{array}{|c|c|c|}
 
\hline 
N=1 ~\textup{superfield} &\sqrt{24}\cdot Y_{U(1)'} & Z_2 \times Z_2' \\
 
\hline \hline
 
V_c, ~V_L, ~V_s, ~V_R & 0 & (+,~+)\\
 
\hline

\Sigma_c,~\Sigma_L,~\Sigma_s,~\Sigma_R & 0 & (-,~-)\\ 

\hline 

V_t,~ V_{\bar{t}}& 4,-4 & (-,~+) \\ 

\hline 

\Sigma_t,~\Sigma_{\bar{t}}& 4,-4 & (+,~-) \\ 

\hline \hline

 q~,~q' & 1 & (+,~+),~(-,~+) \\ 

\hline 

 l~,~l' & -3 & (-,~+),~(+,~+) \\ 

\hline 

 q^c~,~{q^c}' & -1 & (+,~+),~(-,~+)\\ 

\hline 

l^c~,~{l^c}' & 3 &(-,~+),~(+,~+) \\

\hline\hline

\Phi & 0 & (+,~+)\\

\hline\hline

l^c_H~,~{l^c_H}' & 3,-3 & (+,~+) \\

\hline

q^c_H~,~{q^c_H}' & -1,1 & (-,~+)\\

\hline

\end{array}$$

\end{table}
The $SU(2)_R\times U(1)' \to U(1)_Y$ breaking occurs through 
$l^c_H$, ${\ov{l}^c_H}'$ states, after their neutral scalar components 
$\nu^c_H$, ${\ov{\nu }^c_H}'$ have developed non-zero VEV's. As for the models
considered above, here we also introduce a discrete ${\cal Z}$
symmetry. The combination ${\ov{l}^c_H}'l^c_H$ transforms similarly to
(\ref{Ssym}) and the relevant couplings will be 
(\ref{supS})-(\ref{potS}) but
with 
${\cal S}$, $\ov{\cal S}$ replaced by $l^c_H$, ${\ov{l}^c_H}'$. Thus, 
$\nu^c_H$ and ${\ov{\nu }^c_H}'$ which are components of $l^c_H$ 
and ${\ov{l}^c_H}'$
resp. will have non-vanishing VEV's (\ref{nuvevdef}),
(\ref{nusol}). 
This provides the symmetry breaking 
down to $G_{321}$ 
at an intermediate mass scale between the SUSY 
scale $m$ and the unification scale $M$. As a first example suppose 
that $M\sim 10^{16}$GeV in which case $v\sim 10^{16} 10^{-7/(n-1)}$GeV. 
With $n=7$ we get $v\stackrel{<}{_\sim} M$, the value needed to 
have successful high scale
unification in the {\bf I'}-susy422 model (see below). On the other hand
if 
the unification scale is as small as $M\sim 10^6$GeV, with $n=2$ we 
obtain $v\sim 10^4$GeV as needed in low scale unification scenarios.

With $\Phi $ transforming under ${\mathcal Z}$ as
\begin{equation}
            \Phi \to e^{i\frac{\pi}{n}}\Phi~,
\end{equation}  
the coupling responsible for the $\mu$-term
generation is 
\begin{equation}
W_{\mu}=M \left(\frac{{\ov{l}^c_H}'l^c_H}{M^2}\right)^{n-1}
\Phi^2,
\end{equation}
and we obtain the same $\mu $ term as in (\ref{muterm}).
The 4D Yukawa superpotential, responsible for generation of up-down quark and
charged
lepton masses, is
\begin{equation}
W_Y=qq^c\Phi +l'{l^c}'\Phi ~.
\end{equation}
{}From this coupling the Dirac type coupling for the neutrinos is also
generated and $\hat{m}^D_{\nu }\sim \hat{m}_e$.
There is therefore no additional suppression of the Dirac neutrino masses.
For a low fundamental scale this turns out to be a problem. To overcome
this
difficulty we can introduce
an additional singlet state ${\mathcal N}$. The relevant couplings are
\begin{equation}
\frac{1}{M} \left(\frac{{l^c_H}'l^c_H}{M^2}\right)^k 
{\mathcal N} l^c_H \Phi l'+
{l^c_H}' {l^c}'  {\mathcal N} .
\la{nu4221model}
\end{equation}
Due to the last term in (\ref{nu4221model}) the state ${\nu^c}'$ decouples at
the scale $\lan \nu^c_H\ran \equiv M_I$ and the first term, 
which violates lepton
number, is suppressed by appropriate powers of $M_I/M$. This can lead to
a neutrino mass generation of the needed magnitude.
It is not difficult to verify that together with the ${\nu^c}'{\nu }'h_u$
coupling, the operators in (\ref{nu4221model}) give 

\beq
m_{\nu }\sim \l \fr{M_I}{M}\r^{2k}\fr{\lan h_u^0\ran^2}{M}~.
\la{numass3221}
\eeq
Note that we consider $M_I\ll M $ and therefore for
$k>0$
one can obtain the needed suppression.
{}For instance, with $k=1, 2$, $M_I/M\sim 3\cdot 10^{-4} - 10^{-2}$ and 
$M=(10^{6}-10^{9})$~GeV, one can have $m_{\nu }\sim 1$~eV.

\subsubsection{Gauge coupling unification in 5D SUSY with 
$G_{422}\to G_{3221}$\\
intermediate breaking}

Here we address the question of gauge coupling unification for
$G_{422}$ with compactification breaking to $G_{3221}$. 

{\bf Model I'-susy422}

The field content of this scenario is minimal: $N=2$ gauge
supermultiplets, scalar superfields, and $\eta $ generation of matter in
the
bulk. All these states, their $U(1)'$ charges, and $Z_2\tm Z_2'$ parities
are presented in
Table \ref{t:LR3}. The corresponding $b^{M_I}$, $\gamma $, and $\delta $
factors
are
\begin{equation}
           (b_{U(1)'},~b_{SU(2)_R},~b_2,~b_3)^{M_I}=\left(\frac{15}{2},~2,~1,-3\right),
\end{equation}
$$
            (\gamma_{U(1)'},~\gamma_{SU(2)_R},~\gamma_2,~\gamma_3)=\left(3,~0,-2,-6\right)+4\eta\left(1,~1,~1,~1\right),
$$
\begin{equation}
           (\delta_{U(1)'},~\delta_{SU(2)_R},~\delta_2,~\delta_3)=\left(-7,~6,~0,~2\right)+4\eta\left(1,~1,~1,~1\right).
\end{equation}
On the scale $M_I$ we have  matching conditions 
(\ref{match}), (\ref{BsupG1G2}), where $G_1=SU(2)_R$, $G_2=U(1)'$ and
$\tan \te =\sq{\fr{3}{2}}$.
Taking all this into account, from (\ref{als})-(\ref{alG}) we obtain
\beq
\alpha_3^{-1}= \frac{12}{7}\alpha_2^{-1}-\frac{5}{7}\alpha_1^{-1}-
\frac{6}{7\pi}\ln{\frac{M_G}{M_I}}-\frac{3}{7\pi}(2S_1-3S_2),
\eeq
\beq
\ln{\frac{M_I}{M_Z}}=
\frac{5\pi}{14}(\alpha_1^{-1}-\alpha_2^{-1})-\frac{4}{7}\ln{\frac{M_G}{M_I}}-
\frac{1}{7}(4S_1+S_2),
\eeq
\beq
\alpha_G^{-1}=\alpha_2^{-1}-\frac{1}{2\pi}\ln{\frac{M_G}{M_Z}}+
\frac{1-2\eta}{\pi}S_1-\frac{2\eta}{\pi}S_2.
\eeq
In Table \ref{t:LR5} we present several solutions of the above
equations. 
The experimental lower bound
 $M_I \stackrel{>}{_\sim } 10^{4}$GeV 
for the $SU(2)_R$ symmetry breaking scale \cite{pdg}, \cite{suRbound} puts
a bound
also on the
unification scale
$M_G \stackrel{>}{_\sim } 10^{11}$GeV.
The picture of unification for the case in the last row of Table
\ref{t:LR5} is presented in Fig. 3.

\begin{table} \caption{Unification in the model {\bf I'}-susy422,
with
$\eta=0$. Mass scales are measured in GeV units and $\alpha_3=0.119$.
In all cases $\al_G\simeq 0.03$ is perturbative.}

\label{t:LR5} $$\begin{array}{|c|c|c|c|c|c|}
 
\hline 
M_G/\mu_0 & (N,~N') & \log_{10}{[M_I]}  &\log_{10}{[\mu_0]}& 
\log_{10}{[M_G]}  \\
 
\hline \hline

50 & (24,~24) & 5.40 & 10.32 & 12.02 \\

\hline

51 & (24,~25) & 5.18 & 10.20 & 11.91 \\

\hline

55 & (26,~27) & 4.31 & 9.75 & 11.49 \\

\hline

56 & (27,~27) & 4.10 & 9.64 & 11.39 \\

\hline

\end{array}$$
 
\end{table}

{\bf Model II'-susy422: low scale unification}

In order to have low scale unification, as for the model {\bf
III}-susy422, we introduce
two ${\bf 6}_{N=2}^{(i)}$  plets of $SU(4)_c$ (i=1, 2) with
decomposition (\ref{6pl}), and
two  ${\bf 2}_N=2^{(j)}=(D, ~\ov{D})^{(j)}$ of $SU(2)_L$ ($j=1, 2$).
In addition we introduce 
one bi-doublet ${\bf \Phi'}_{N=2}$ state of (\ref{bidoublets}). 
We will take $Z_2\tm Z_2'$ parities (\ref{trippar}) and (\ref{2par}) for
fragments
from
${\bf 6}_{N=2}^{(i)}$ and ${\bf 2}_{N=2}^{(j)}$, resp.,
while for the fragments of
${\bf \Phi'}_{N=2}$ we take the
parities $\Phi'\sim(-,+)$, $\bar{\Phi}'\sim(+,-)$. The contributions of
these states to the $b$, $\gamma$, and $\delta$ factors are
\begin{equation}
           (b_{U(1)'},~b_{SU(2)_R},~b_2,~b_3)^{T}=\left(1,~0,~0,~1\right),
\end{equation}
\begin{equation}
            (b_{U(1)'},~b_{SU(2)_R},~b_2,~b_3)^D=\left(0,~0,~1,~0\right),
\end{equation}
$$
           ({\bf \Delta_6}+{\bf \Delta_2})(\gamma_{U(1)'},~\gamma_{SU(2)_R},~\gamma_2,~\gamma_3)=\left(2,~0,~0,~2\right)+\left(0,~0,~2,~0\right),
$$
\begin{equation}
           ({\bf \Delta_6}+{\bf \Delta_{\Phi'}})(\delta_{U(1)'},~\delta_{SU(2)_R},~\delta_2,~\delta_3)=\left(2,~0,~0,~2\right)+\left(0,~2,~2,~0\right).
\end{equation}
With these changes we obtain
\begin{equation}
\alpha_3^{-1}=\frac{12}{7}\alpha_2^{-1} -\frac{5}{7}\alpha_1^{-1}-
\frac{6}{7\pi}\ln{\frac{M_G}{M_I}}+\frac{9}{14\pi}\ln{\frac{M_G}{M_T}}-
\frac{6}{7\pi}\ln{\frac{M_G}{M_D}} 
 -\frac{9}{7\pi}\left(S_1 - S_2\right),
\la{al3}
\end{equation}
\begin{equation}
\ln{\frac{M_I}{M_Z}}=\frac{5\pi}{14}(\alpha_1^{-1}-\alpha_2^{-1})- 
\frac{4}{7}\ln{\frac{M_G}{M_I}}-\frac{1}{14}\ln{\frac{M_G}{M_T}}+
\frac{5}{28}\ln{\frac{M_G}{M_D}}
-\frac{1}{14}\left(5\,S_1+2\,S_2\right),
\la{scaleMI}
\end{equation}
\begin{equation}
\alpha_G^{-1}=\alpha_2^{-1}-\frac{1}{2\pi}\ln{\frac{M_G}{M_Z}}-
\frac{1}{2\pi}\ln{\frac{M_G}{M_D}}-\frac{2\eta}{\pi}S_1-
\frac{1+2\eta}{\pi}S_2.
\end{equation}
In (\ref{al3}) the last term vanishes in the $S_1=S_2$ limit, 
while the last term of
(\ref{scaleMI}) is negative. This shows that low scale unification is possible.
In Table \ref{t:LR6} we present several cases of successful unification in
this scenario. With the experimental bound 
$M_I \stackrel{>}{_\sim } 10^4$GeV we
have $M_G \stackrel{>}{_\sim } 10^6$GeV. 
The picture of unification for the $M_G/\mu_0=110$ case of Table
\ref{t:LR6} is presented in Fig. 4.

\begin{table} \caption{Unification in the model {\bf II'}-susy422, with
$\eta=0$. Mass scales are measured in GeV units and $\alpha_3=0.119$.
In all cases $\al_G\simeq 0.1$ is perturbative.}

\label{t:LR6} $$\begin{array}{|c|c|c|c|c|c|c|c|}
 
\hline 
M_G/\mu_0 & (N,~N') & \log_{10}{[M_G/M_T]} & \log_{10}{[M_G/M_D]} 
& \log_{10}{[M_I]}  &\log_{10}{[\mu_0]}& \log_{10}{[M_G]} \\
 
\hline \hline

110 & (54,~54) & 3.44 & 2.52 & 3.85  & 3.86 & 5.90 \\

\hline

109 & (53,~54) & 3.62 & 2.65 & 3.97 & 3.98 & 6.00 \\

\hline

108 & (53,~53) & 3.80 & 2.80 & 4.10 & 4.10 & 6.15 \\

\hline

107 & (52,~53) & 3.78 & 2.78 & 4.21 & 4.21 & 6.23 \\

\hline

90 & (44,~44) & 5.26 & 1.96 & 4.71 & 6.67 & 8.63 \\

\hline

89 & (43,~44) & 5.42 & 1.95 & 4.73 & 6.82 & 8.77 \\

\hline

80 & (39,~39) & 7.90 & 1.90 & 4.45 & 8.46 & 10.36 \\

\hline

\end{array}$$
 
\end{table}

{\bf Model III'-susy422: low scale unification}

A different model which also gives low scale unification is a scenario
extended with ${\bf \Phi'}_{N=2}$ and four
${\bf R}^{r}_{N=2}=(R ,~\ov{R})^{(r)}$ ($r=1,\cdots ,4$), where 
$R$ is a doublet of $SU(2)_R$.
With $Z_2\tm Z_2'$ parities
\begin{equation}
R^{(r)} \sim(+,~+)~,~~~\ov{R}^{(r)}\sim (-, ~-)~,
\end{equation}
their contributions to the $b$-factors (above the common mass $M_R$) and 
$\gamma $, $\delta $ factors are
\begin{equation}
(b_{U(1)'},~b_{SU(2)_R},~b_2,~b_3)^{M_R}=\left(0,~2,~0,~0\right),
\end{equation}
\begin{equation}
                      {\bf \Delta_R}(\gamma_{U(1)'},~\gamma_{SU(2)_R},~\gamma_2,~\gamma_3)=(0,~4,~0,~0).
\end{equation}
RGEs in this case are
\begin{equation}
                     \alpha_3^{-1} = \frac{12}{7}\alpha_2^{-1}  -\frac{5}{7}\alpha_1^{-1} - \frac{6}{7\pi}\ln{\frac{M_G}{M_I}} + \frac{3}{7\pi}\ln{\frac{M_G}{M_R}},
\end{equation}
\begin{equation}
                     \ln{\frac{M_I}{M_Z}}= \frac{5\pi}{14}(\alpha_1^{-1} - \alpha_2^{-1}) - \frac{4}{7}\ln{\frac{M_G}{M_I}} - \frac{3}{14}\ln{\frac{M_G}{M_R}} - S_1,
\end{equation}
\begin{equation}
                     \alpha_G^{-1}=\alpha_2^{-1}-\frac{1}{2\pi}\ln{\frac{M_G}{M_Z}} + \frac{1-2\eta}{\pi}S_1-\frac{1+2\eta}{\pi}S_2.
\end{equation}
{}From the above equations we see that this scenario also allows low scale
unification. In Table
\ref{t:LR7} we present mass scales giving successful unification.
In this case the experimental bound 
$M_I\stackrel{>}{_\sim }10^4$~GeV gives  
$M_G \stackrel{>}{_\sim }10^{5.8}$GeV. 
The picture of unification for the $M_G/\mu_0=57$ case of Table
\ref{t:LR7} is presented in Fig. 5.

\begin{table} \caption{Unification in the model {\bf III'}-susy422, with
$\eta=0$. Mass scales are measured in GeV units and $\alpha_3=0.119$.
In all cases $\al_G\simeq 0.04$ is perturbative.}

\label{t:LR7} $$\begin{array}{|c|c|c|c|c|c|c|}
 
\hline 
M_G/\mu_0 & (N,~N') & \log_{10}{[M_G/M_R]} & \log_{10}{[M_I]} 
&\log_{10}{[\mu_0]}&\log_{10}{[M_G]}  \\
 
\hline \hline

57 & (27,~28) & 2.89 & 3.41 & 3.41 & 5.17 \\

\hline

55 & (26,~27) & 2.85 & 3.85 & 3.85 & 5.59 \\

\hline

54 & (26,~26) & 2.84 & 4.08 & 4.08 & 5.81 \\

\hline

53 & (25,~26) & 2.82 & 4.30 & 4.30 & 6.02 \\

\hline

37 & (17,~18) & 7.11 & 5.55 & 7.85 & 9.42 \\

\hline

36 & (17,~17) & 7.29 & 5.67 & 8.07 & 9.63 \\

\hline

18 & (8,~8) & 11.08 & 7.54 & 12.14 & 13.39 \\

\hline

17 & (7,~8) & 11.24 & 7.66 & 12.36 & 13.59 \\

\hline

\end{array}$$
 
\end{table} 

\locsection{5D flipped $SU(5)\times U(1)$ model \\
on $S^{(1)}/Z_2\times Z_2'$ orbifold}

Finally we consider flipped $SU(5)\times U(1)\equiv G_{51}$ GUT in 5D.
The decomposition of $SU(5)$'s adjoint $V(24)$ in terms of
$SU(3)_c\times SU(2)_L\times U(1)'$ reads
\beq
V(24)=V_c(8, 1)_0+V_{SU(2)_L}(1, 3)_0+V_s(1, 1)_0+V_X(3, \bar 2)_5+
V_Y(\bar 3, 2)_{-5}~,
\la{dec24of51}
\eeq
where subscripts denote $U(1)'$ charges in $1/\sq{60}$ units. Since
the $U(1)'$ comes from $SU(5)$, it has the usual $1/\sq{60}$
normalization.
%
$V_s$ in (\ref{dec24of51}) corresponds to the $U(1)'$ gauge
superfield. The
decomposition of $\Si (24)$ has an identical form.

{\bf Matter ~sector}

The
matter sector contains $\eta $ generations of $N=2$ supermultiplets
\beq
{\bf X}_{N=2}=({\cal X},~\ov{\cal X})~,~~~~
{\bf V}_{N=2}=({\cal V},~\ov{\cal V})~,~~~~
{\bf I}_{N=2}=({\cal I},~\ov{\cal I})~,
\la{matter51}
\eeq
where
\beq
{\cal X}={\bf 10}_{-1}=(\nu^c ,~q ,~d^c)_{-1}~,~~
{\cal V}=\bf{\bar 5}_3=(l ,~u^c)_3~,~~
{\cal I}={\bf 1}_{-5}=e^c_{-5}~
\la{matterdec51}
\eeq
are the usual chiral multiplets of flipped $SU(5)\times U(1)$ GUT, and
$\ov{\cal X}$, $\ov{\cal V}$, $\ov{\cal I}$ in (\ref{matter51}) are
their mirrors.
Subscripts in (\ref{matterdec51}) denote $U(1)$ charges, defined up to
some normalization factor. Assuming that $G_{51}$ comes from $SO(10)$, the
normalization factor will be $1/\sq{40}$. 
Therefore, in $SO(10)$ normalization
\beq
Y_{U(1)}[{\cal X}]=-\fr{1}{\sq{40}}~,~~~
Y_{U(1)}[{\cal V}]=\fr{3}{\sq{40}}~,~~~
Y_{U(1)}[{\cal I}]=-\fr{5}{\sq{40}}~,~
\la{so10norm}
\eeq
In fact, the $SO(10)$ spinor ${\bf 16}$ in terms of $G_{51}$ reads
\beq
{\bf 16}_{SO(10)}={\bf 10}_{-1}+{\bf \bar 5}_{3}+{\bf 1}_{-5}~.
\la{decso10}
\eeq  

We also introduce $\eta $ copies ${\bf X'}_{N=2}$, ${\bf V'}_{N=2}$,
${\bf I'}_{N=2}$. $3-\eta $ generations are introduced at the brane.

{\bf Scalar sector}

To have MSSM doublets we introduce the following set of supermultiplets:
\beq
{\bf h}_{N=2}=\l h(5_2) ,~\ov{h}(\bar 5_{-2})\r ~,~~~~
{\bf h}_{N=2}'=\l h'(5_{-2}) ,~\bar h'(\bar 5_{-2})\r ~,
\la{N2sc51doubl}
\eeq
where subscripts denote $U(1)$ charges in $1/\sq{40}$ units. 
In (\ref{N2sc51doubl}) one has
\beq
h(5_2)=(h_d ,~\ov{d}^c_h)_{2}~,~~~~~
\ov{h}(\bar 5_{-2})=(h_u ,~d^c_{\bar h})_{-2}~,
\la{N1sc51doubl}
\eeq 
and the same for states with primes. 

For reducing the rank of the group via Higgs breaking we need additional
states. Thus we
introduce
\beq
{\bf H}_{N=2}=\l H , ~\ov{H} \r~,~~~~~
{\bf H}_{N=2}'=\l H' ,~\ov{H}\hs{0.03cm}'\r~,
\la{N2H51}
\eeq
where
\beq
H=10_{-1}^H=\l \nu^c_H ,~d^c_H ,~q^{~}_H \r_{-1}~,~~~
\ov{H}=\ov{10}^{H}_1=\l \ov{\nu }^c_H ,~\ov{d}^c_H
,~\ov{q}^{~}_H\r_{1}~,
\la{N1H51}
\eeq
and similarly for $H'$ and $\ov{H}\hs{0.03cm}'$.

\subsection{$G_{51}\to G_{321}$ breaking and related phenomenological
questions}

The first stage of breakdown of the $G_{51}$ gauge group again occurs through
orbifolding: prescribing to various fragments of $G_{51}$
multiplets certain $Z_2\times Z_2'$ parities some states are projected
out, and at a fixed point we remain with a reduced gauge group. The
transformation properties of the fragments of 5D SUSY $G_{51}$ are given
in Table \ref{t:51GUT}.
%
%
%
%
\begin{table} \caption{$U(1)$, $U(1)'$ charges and $Z_2\times Z_2'$
parities of various fragments in 5D SUSY $G_{51}$ scenario. All mirrors,
of matter and scalars, have opposite charges and parities.}
 
\label{t:51GUT} 

$$\begin{array}{|c|c|c|c|}
 
\hline N=1~{\rm superfield} &\sq{40}\cdot Y_{U(1)} & 
\sq{60}\cdot Y_{U(1)'} & Z_2\times Z_2' \\
 
\hline \hline
 
V_c,~V_{SU(2)_L},~V_{U(1)},~V_s & 0 &0 & (+,~+) \\
 
\hline
 
\Si_c,~\Si_{SU(2)_L},~\Si_{U(1)},~\Si_s&0 &0 &(-,~-) \\
 
\hline
 
V_X~,~V_Y& 0 &5 ,~-5  &(- ,~+ ) \\
 
\hline
 
\Si_X~,~\Si_Y& 0 &5 ,~-5   &(+ ,~- ) \\
 
\hline \hline
 
q~~,~~q\hs{0.6mm}'&-1   &-1  &(+ ,~+ ),~(-,~+) \\
 
\hline
 
l~~,~~l\hs{0.6mm}'& 3 &3  &(+ ,~+ ),~(-,~+) \\
 
\hline
 
u^c~~,~~{u^c}'& 3 &-2  &(- ,~+ ),~(+,~+) \\     

\hline
 
d^c~~,~~{d^c}'&-1 &4  &(- ,~+ ),~(+,~+) \\
 
\hline
 
\nu^c~~,~~{\nu^c}'& -1 &-6  &(- ,~+ ),~(+,~+) \\
 
\hline
 
e^c~~,~~{e^c}'& -5 &0  &(+ ,~+ ),~(-,~+) \\
 
\hline \hline

h_u~~,~~{h_d}'&-2 ,~ 2  &3 ,~ -3   &(+ ,~+ ) \\
 
\hline
 
{d^c}_{\bar h}~~,~~
{\bar{d}^{c'}_h}&-2 ,~ 2  &-2 ,~ 2  &(- ,~+ ) \\
 
\hline              

\nu^c_H~~,~~{\ov{\nu }^c_H}'& -1 ,~ 1  &-6 ,~ 6  &(+ ,~+ ) \\
 
\hline
 
d^c_H~~,~~{\ov{d}^c_H}'&-1 ,~ 1   &4 ,~ -4   &(+ ,~+ ) \\
 
\hline

q^{~}_H~~,~~{\ov{q}^{~}_H}'&-1 ,~ 1   &-1 ,~ 1   &(- ,~+ ) \\
 
 
 
 
 
\hline

\end{array}$$
 
\end{table} 
%
%
%
%
One can easily see that at the $y=0$ fixed point
we have a $SU(3)_c\tm SU(2)_L\tm U(1)\tm U(1)'\equiv {G'}_{3211} $ gauge
group. At this fixed point together with three generations of
quark-lepton and right handed neutrino superfields
$q$, ${u^c}'$, ${d^c}'$, $l$, $e^c$, ${\nu^c}'$ and an MSSM pair of Higgs
doublets 
$h_u$, ${h_d}'$, we also have the states
$d^c_H$, ${\ov{d}^c_H}'$, $\nu^c_H$, ${\ov{\nu }^c_H}'$. The
latter two states are responsible for an $U(1)\tm U(1)'$ breaking down to
$U(1)_Y$. If they develop VEVs on the scale $M_I$ the second stage
of symmetry breaking occurs and we have the unbroken $U(1)_Y$ generator
\beq
Y=\fr{2\sq{6}}{5}Y_{U(1)}-\fr{1}{5}Y_{U(1)'}~,
\la{sup51charges}
\eeq
where $Y_{U(1)'}$ and $Y_{U(1)}$
charges
of the states are given in Table \ref{t:51GUT}. [For $Y_{U(1)}$
again we have used $SO(10)$ normalization (\ref{so10norm}), since flipped
$SU(5)\tm U(1)$
is one of its maximal subgroups \cite{SU51ofSO10}, \cite{slansky}].

The generation of $\lan \nu^c_H\ran $, $\lan {\ov{\nu }^c_H}'\ran $ VEVs can
happen in the same way as for the $SU(4)_c\tm SU(2)_L\tm SU(2)_R$ scenario
presented in sect. 7.1. 
Introducing a discrete ${\cal Z}$ symmetry and a transformation for 
${\ov{\nu }^c_H}'\nu^c_H$ as in
(\ref{Ssym}), the relevant superpotential, soft SUSY breaking,
and all the potential
couplings
will have the form of 
(\ref{supS}), (\ref{softSpot}), (\ref{potS}) resp. if the states 
${\cal S}$, $\ov{\cal S}$ are replaced by 
$\nu^c_H$, ${\ov{\nu }^c_H}'$. 
{}From all this one can ensure non zero VEV solutions
(\ref{nuvevdef}), (\ref{nusol}). 
By a proper choice of $n$ one can obtain a gap between the mass
scales $v\equiv M_I$ and $M$.
Also $\mu $ term generation can happen similarly. With
transformation (\ref{mutermMSSM}) for the $h_u{h_d}'$ combination and
coupling (\ref{musup})  the
$\mu $ term (\ref{muterm}) is derived.

%

Also for the SUSY $G_{51}$
model we use a
${\cal Z}$ symmetry to obtain a realistic phenomenology.
The Lagrangian (\ref{lagv}), (\ref{lagf}) with $Z_2\times
Z_2'$ orbifold
parities should be invariant under this ${\cal Z}$;
therefore the fragments of matter and
scalar superfields from the same $SU(5)\times U(1)$ multiplets should have
identical ${\cal Z}$ phases, while the phases of mirrors must
be opposite.  Together with quark-lepton superfields and
right handed neutrinos at the $y=0$ fixed point we have zero mode $\nu^c_H$,
${\ov{\nu }^c_H}'$, $d^c_H$, ${\ov{d}^c_H}'$ states.
{}With couplings (\ref{supS}) 
(with $\ov{\cal S}{\cal S}$ replaced by ${\ov{\nu }^c_H}'{\nu }^c_H$), 
(\ref{musup}) we thus have
\beq
\al (\nu^c_H)+\al ({\ov{\nu }^c_H}')=
\al ({h_d}')+\al (h_u)=
\fr{2\pi }{n}~,
\la{phasecond1}
\eeq
and 
\beq
\al ({\nu^c}')=\al ({d^c}')~,~~~
\al (\nu^c_H)=\al (d^c_H)~,~~~
\al ({\ov{\nu }^c_H}')=\al ({\ov{d}^c_H}')~.
\la{phasecond2}
\eeq
Therefore, at this stage together with the phases of $q$, $u^c$, $l$,
$e^c$,
we have 8 independent phases.
Writing Yukawa couplings which generate masses for up-down quarks and
charged leptons
\beq
W_Y=q{u^c}'h_u+q{d^c}'{h_d}'+le^c{h_d}'~,
\la{yuk51}
\eeq
we will remain with 4 independent phases. It is possible to fix two of
them 
from the neutrino sector, writing Dirac and Majorana type couplings. For the question of
the mass
scales needed for successful unification the
couplings of type (\ref{dirmaj422}) are
irrelevant, while terms
${\nu^c}'lh_u+\fr{1}{M}({\ov{\nu }^c_H}'{\nu^c}')^2$ give
$m_{\nu }\sim \fr{h_u^2}{M_I^2}M$. The latter leads to the reasonable
value 
$m_{\nu }\sim 1$~eV for $M_G/\mu_0=1$ in Table
\ref{t:51susy}. However,
through these couplings the phases of appropriate states are selected in
such a way, that some unacceptably large matter parity (lepton
number) and baryon number violating couplings are allowed. To
avoid this, we will modify to a model, which gives a value
for the neutrino mass accommodating atmospheric data and does not
lead to
any unacceptable process. Including a matter parity violating
(\ref{bilinS}) type coupling
\beq
\l \fr{{\ov{\nu }^c_H}'\nu^c_H}{M^2}\r^{n-1}\nu^c_Hlh_u~,
\la{bilin}
\eeq
similar to the treatment in the MSSM and $SU(5)$ cases,
one obtains an operator 
$\mu_l lh_u$ with $\mu_l\sim \fr{M_I}{M}m$
inducing a neutrino mass (\ref{nubilinS}),
where in this case
$\sin \xi \sim \fr{M_I}{M}$
(still without assuming
any alignment between
superpotential and soft SUSY breaking couplings). 
For a suppressed neutrino mass we
need $M_I/M\leq 3\cdot 10^{-6}$. This
case indeed is realized with successful unification [see case 
$M_G/\mu =5$
in
Table \ref{t:51susy}]: for $M_I\simeq 6\cdot 10^{10}$~GeV and
$M=M_G\simeq 2\cdot 10^{16}$~GeV, three gauge
couplings are unified and we have $m_{\nu }\stackrel{<}{_\sim } 1$~eV.

Couplings (\ref{yuk51}), (\ref{bilin}) and conditions
(\ref{phasecond1}), (\ref{phasecond2})  determine
the phases
$$
\al ({h_d}')=\al -\al (h_u)~,~~\al ({u^c}')=-\al (q)-\al (h_u)~,~~
\al (e^c)=-\al (l)+\al (h_u)-\al ~,
$$
$$
\al ({\nu^c}')=\al ({d^c}')=-\al (q)+\al (h_u)-\al ~,~~
\al (\nu^c_H)=\al(d^c_H)=\al -\al (l)-\al (h_u)~,
$$
\beq
\al ({\ov{\nu }^c_H}')=\al({\ov{d}^c_H}')=\al (l)+\al (h_u)~,~~~
\al =\fr{2\pi }{n}~,
\la{phases51}
\eeq
but $\al (q)$, $\al (l)$, $\al (h_u)$ are still
free. With eqs. (\ref{phases51}) the lepton number
violating operators
\beq
\frac{\nu^c_H}{M}q{d^c}'l~,~~~~~
\frac{\nu^c_H}{M}e^cll~,
\la{trilinlep}
\eeq
are allowed, and for $M_I/M\sim 3\cdot 10^{-6}$ they lead to a radiative
neutrino
mass $m_{\nu }'\sim 10^{-5}$~eV [see eq. (\ref{numastrilin}), 
(\ref{trilinS1})], with the relevant scale for the solar neutrino anomaly.

Chosing $\al (q)$, $\al (l)$  and $\al (h_u)$ as 
\beq
\al (q)=\al (l)=\al /3,~~~~~\al (h_u)=\al ~,
\la{phaseqlh}
\eeq
the discrete ${\cal Z}$ symmetry will be $Z_{3n}$. 
With assignment (\ref{phases51}), (\ref{phaseqlh}) the coupling
$\nu^c_Hqld^c_H$ is allowed. However, the coupling 
${\ov{\nu }^c_H}'qq{\ov{d}^c_H}'$ is forbidden and therefore the $d=5$
operator $qqql$ does not emerge. 
One can  verify easily that 
all other matter parity and
baryon number violating operators are absent in this setting.
The coupling $\nu^c_Hqld^c_H$ would induce decays of a light triplet 
$d^c_H$ (mass $\sim $ few TeV) $d^c_H\to ql$ with leptoquark signature,
observable in future collider experiments \cite{leptq}.

Concluding this subsection we note that in order to give masses to the
states ${\nu^c}'$ one can introduce ${\cal N}$ singlet states with
$Z_{3n}$ phase $\al ({\cal N})=-\al $. Through couplings 
${\ov{\nu}^c_H}'{\nu^c}'{\cal N}$ right handed states would decouple at
the scale $M_I$. 
%

\subsection{Gauge coupling unification in 5D SUSY $G_{51}$}

Below $M_I$ we have the MSSM field content with the b-factors 
(\ref{Bmssm}),
plus
states $d^c_H$, ${\ov{d}^c_H}'$ with mass $M_{d^c_H}$ in the TeV range
and  b-factors as in (\ref{Bdc}).
Above the $M_I$ scale we obtain
\beq 
\l b_{U(1)},~b_{U(1)'},~b_2,~b_3\r^{M_I} = 
\l \fr{33}{5},~\fr{47}{5},~1,~-2\r~,
\la{51B3211} 
\eeq
and for the $\ga $ and $\de $ factors 
$$
\l \ga^{~}_{U(1)},~\ga^{~}_{U(1)'},~\ga_2,~\ga_3\r =
\l \fr{6}{5},~\fr{34}{5},~-2,~-4\r+
4\eta (1,~1,~1,~1)~,
$$
\beq 
\l \de_{U(1)},~\de_{U(1)'},~\de_2,~\de_3\r =
\l \fr{9}{5},~-\fr{44}{5},~0,~2 \r+
4\eta (1,~1,~1,~1)~,
\la{BKK51}
\eeq
if $\eta $ families have KK excitations.

According to (\ref{sup51charges}), we get  
$\tan \te=2\sq{6} $ for the $\tan \te $ in (\ref{match}) if $G_1=U(1)$ and
$G_2=U(1)'$. Taking into account
this and
also (\ref{RGE321})-(\ref{alG}) one obtains
\beq 
\al_3^{-1}=\fr{12}{7}\al_2^{-1}-\fr{5}{7}\al_1^{-1}+
\fr{27}{50\pi }\ln \fr{M_G}{M_I}+
\fr{9}{14\pi }\ln \fr{M_I}{M_{d^c_H}}+ 
\fr{39}{175\pi }S_1+
\fr{261}{175\pi}S_2~,
\la{51modelals} 
\eeq 
\beq 
\ln \fr{M_I}{M_Z}=\fr{5\pi }{14}(\al_1^{-1}-\al_2^{-1})- 
\fr{51}{50}\ln \fr{M_G}{M_I}-
\fr{1}{14}\ln \fr{M_I}{M_{d^c_H}}-
\fr{107}{175}S_1-\fr{43}{175}S_2~, 
\la{51modelscale} 
\eeq 
\beq 
\al_G^{-1}=\al_2^{-1}-\fr{1}{2\pi }\ln \fr{M_G}{M_Z}+
\fr{1-2\eta }{\pi }S_1-\fr{2\eta }{\pi }S_2~.
\la{51modelalG}  
\eeq 
The last four terms in (\ref{51modelals}) are positive and in order to
get a
reasonable value for $\al_3$ one should take 
$\mu_0\simeq M_G\simeq M_I\simeq M_{d^c_H}$. Then from 
(\ref{51modelscale}) one obtains $M_I\sim 10^{16}$~GeV ($\simeq
M_{d^c_H}$). On the other hand we know that this scenario leads to 
light
$d^c_H$, ${\ov{d}^c_H}'$ states, inconsistent with 
unification. To resolve this problem we can introduce additional $N=2$
SUSY
states, which contain zero mode $SU(2)_L$ doublets, which being
light ($\sim $TeV) will compensate contributions from colored triplets:
states 
${\bf \Psi}^{(i)}_{N=2}=(\Psi,~\ov{\Psi })^{(i)}$
($i=1, 2$), with $SU(5)\times U(1)$ representations  
$\Psi^{(i)}\sim 5_2 $,
$\ov{\Psi }^{(i)}\sim \bar 5_{-2}$; They decompose into doublets and
triplets 
$\Psi^{(i)}=(D, T)^{(i)}$, $\ov{\Psi }^{(i)}=(\ov{D}, \ov{T})^{(i)}$. With
$Z_2\times Z_2'$ orbifold parities
$$
(D^{(1)},~\ov{D}^{(2)})\sim (+,~+)~,~~~
(D^{(2)},~\ov{D}^{(1)})\sim (-,~-)~,
$$
\beq
(T^{(1)},~\ov{T}^{(2)}))\sim (-,~+)~,~~~
(T^{(2)}, ~\ov{T}^{(1)})\sim (+,~-)~,
\la{DTorbpar}
\eeq 
only $D^{(1)}$,  $\ov{D}^{(2)}$ states will have zero modes and contribute
in the b-factors. Below $M_I$
\beq
\De_{\Psi }\l b_1,~b_2,~b_3\r =\l \fr{3}{5},~1,~0\r ~,
\la{lowbPsi}
\eeq
while above the $M_I$ scale
\beq
\De_{\Psi }\l b_{U(1)},~b_{U(1)'},~b_2,~b_3\r^{M_I}=
\l \fr{2}{5},~\fr{3}{5},~1,~0\r~.
\la{highbPsi}
\eeq
Contributions to the $\ga $ and $\de $ factors from the fragments of 
${\bf \Psi}_{N=2}^{(i)}$ states are
$$
\De_{\Psi }\l \ga^{~}_{U(1)},~\ga^{~}_{U(1)'},~\ga_2,~\ga_3\r =
\l \fr{4}{5},~\fr{6}{5},~2,~0\r ~,~
$$
\beq
\De_{\Psi}\l \de_{U(1)},~\de_{U(1)'},~\de_2,~\de_3\r =
\l \fr{6}{5},~\fr{4}{5},~0,~2\r ~.
\la{gadePsi}
\eeq
Taking into account (\ref{51B3211}), (\ref{BKK51}), 
(\ref{lowbPsi})-(\ref{gadePsi}), we finally have
\beq
\al_3^{-1}=\fr{12}{7}\al_2^{-1}-\fr{5}{7}\al_1^{-1}-
\fr{6}{35\pi }\ln \fr{M_G}{M_I}+  
\fr{9}{14\pi }\ln \fr{M_I}{M_{d^c_H}}-
\fr{9}{14\pi }\ln \fr{M_I}{M_D}-
\fr{6}{5\pi }S_1+
\fr{102}{35\pi}S_2~,
\la{ext51modelals} 
\eeq
\beq
\ln \fr{M_I}{M_Z}=\fr{5\pi }{14}(\al_1^{-1}-\al_2^{-1})-
\fr{32}{35}\ln \fr{M_G}{M_I}- 
\fr{1}{14}\ln \fr{M_I}{M_{d^c_H}}+
\fr{1}{14}\ln \fr{M_I}{M_D}
-\fr{14}{35}S_1-\fr{16}{35}S_2~, 
\la{ext51modelscale}
\eeq
\beq
\al_G^{-1}=\al_2^{-1}-\fr{1}{2\pi }\ln \fr{M_G}{M_Z}-
\fr{1}{2\pi }\ln \fr{M_G}{M_D}-
\fr{2\eta }{\pi }S_1-\fr{2\eta }{\pi }S_2~,
\la{ext51modelalG} 
\eeq
where $M_D$ is the mass of $D^{(1)}$, $\ov{D}^{(2)}$ state's zero modes.
The mass scales, for which successful unification takes place in this
model, are presented in Table \ref{t:51susy}. As we see, the masses of the
doublets are in a range  $1.7$~TeV--$141$~GeV.
The picture of unification for the case in the last row of Table
\ref{t:51susy} is presented in Fig. 6.

\begin{table} \caption{Unification for SUSY $G_{51}$ model. 
(Scales are measured in units of GeV.)}
 
\label{t:51susy} $$\begin{array}{|c|c|c|c|c|c|c|c|}
 
\hline 
M_G/\mu_0&(N ,~N') & \log_{10}[M_I] & 
\log_{10}[\mu_0] & \log_{10}[M_G]& 
\log_{10}[\fr{M_I}{M_{d^c_H}}] &
\log_{10}[\fr{M_I}{M_D}] &\al_3(M_Z)    \\
 
\hline \hline
 
1 &(0 ,~0 ) &15.37   &16.37   &16.37  &12 &12.15 &0.119 \\
 
\hline

3 &(0 ,~1 ) &14.59   &15.61   &16.09  &11 &12.2 &0.1184 \\  

\hline

4 &(1 ,~1 ) &12.79   &15.69   &16.29  &8 &10.2 &0.119\\  

\hline

5&(1 ,~2 ) &10.77 &15.58 &16.27 &6.7 &8.6 &0.117 \\

\hline

6 &(2 ,~2 ) &5.53   &15.75   &16.53  &2 &3.3 &0.1176\\  

\hline

6 &(2 ,~2 ) &6.45   &15.67   &16.45  &3 &4.3 &0.116\\  

\hline

\end{array}$$
 
\end{table} 

Concluding this subsection,  we note that in the flipped $SU(5)\tm U(1)$
GUT it
is impossible to get low scale (near few or multi TeV) unification. The
reason is the following: introducing some additional states, one should
cancel
the positive power law contribution in (\ref{51modelals}) in order to get
a reasonable $\al_3 (M_Z)$. The contribution from additional states 
will have the form 
$\fr{12}{7}{\De '}_2-\fr{5}{7}{\De '}_1-{\De '}_3$, where ${\De '}_a$ is
a contribution to the renormalization of $\al^{-1}_a$. Since 
fragments from non trivial $SU(5)$ representations give the same
contribution
to factors of $SU(2)_L$ and $SU(3)_c$ we have 
${\De '}_3={\De '}_2$. Therefore the final contribution to 
$\al_3^{-1}$ is $\fr{5}{7}({\De '}_2-{\De '}_1)$.
The latter should cancel the last two terms in (\ref{51modelals}), which
in the
$S_1=S_2\equiv S$ limit are equal to $\fr{12}{7\pi }S$. Thus, from the
cancellation condition we have ${\De '}_1-{\De '}_2=\fr{12}{5\pi }S$; 
and thus the contribution 
in (\ref{51modelscale}) 
is $\fr{6}{7}S$. This value precisely cancels the last
two terms of (\ref{51modelscale}) in the $S_1=S_2=S$ limit. This means that
$M_G$ can not be lowered down to multi TeV.  
In Fig. 5 all couplings unify at one point. However, it would be quite
natural in spirit of a two step unification that at a first step only the
three
couplings of $SU(5)$ unify and the $U(1)$ coupling joins at a higher
scale. This can be achieved either by a change of the intermediate scale
or by a different choice of the extra state's masses.

\section{Conclusions and outlook}

We have considered 5D orbifold SUSY models and within them we have
addressed
numerous phenomenological issues. Orbifold constructions give 
an attractive
resolution of several outstanding problems of GUTs, but some
extensions are still needed to have full control of difficulties
which even appear outside 
GUTs. In fact, problems such as baryon number violation, $\mu $
problem and neutrino oscillations (tied with lepton number violation)
are not cured by extra dimensions, and
some care is needed to deal with them. In our
approach we have considered extensions with a discrete ${\cal Z}$
symmetry, which gives a natural and simultaneous understanding of these
problems. In the essential part of the paper we have addressed the
question
of gauge coupling unification,
which in the presence of KK states gets new facets. Since the orbifold
approach to the celebrated $SU(5)$
GUT does not allow for low or intermediate scale unification, we 
studied extended (in rank) GUTs. As we have seen 
this opens up new and interesting possibilities from the viewpoints
of unification and phenomenological implications.
For extended GUTs, symmetry breaking can occur by a step by step
compactification and if the intermediate gauge group differs 
from semisimple $SU(5)$, 
power law unification can take place. As an example 
we have considered the two
maximal subgroups of $SO(10)$ - the Pati-Salam $G_{422}$ and flipped
$G_{51}$ GUTs. Within $G_{422}$, low scale unification can take place,
while $G_{51}$ only allows for unification at scales 
$\sim 10^{16}$~GeV. 
The latter scales are also possible for $G_{422}$.
Within both scenarios extensions with a
discrete ${\cal Z}$ symmetry were pursued; thus key phenomenological
problems
were resolved in an elegant way. The $G_{51}$ model and also in some cases
the
$G_{422}$ one
predict colored triplet states in the few TeV range.
$G_{422}$ models with low scale unification lead to additional
relatively light states. Because of these the models  have an
interesting phenomenology, in particular if KK states appear near the TeV
scale.
Future high energy collider
experiments will test the relevance of such models.

In our studies we have used one loop RGE analysis. Two loop 
power law contributions to the 
$\beta $-function would contribute  significantly. This would
make estimates unstable. However, due to higher dimensional
supersymmetries, in higher loops  full sets of $N=2$
supermultiplets can contribute \cite{lowunif1}, \cite{lowunif2} and in
this case effects of higher loops will be
logarithmic. The latter contributions would not change the unification
picture and might only imply a slight modification of mass scales.
The same argument could be applied for threshold corrections coming from
different sources.

Our studies of the $G_{422}$ and $G_{51}$ groups were performed in a
bottom-up approach and therefore the 'low energy' sector was more
relevant for our considerations. Having in mind that both $G_{422}$ and
$G_{51}$
are maximal subgroups of $SO(10)$, it would be interesting to construct
a higher (at least two extra) dimensional $SO(10)$ model 
where extra dimensions do not compactify at a single 
mass scale (in contrast to the models of \cite{orbSO10})
but with  intermediate groups
$G_{422}$ or $G_{51}$ at a scale between $\fr{1}{R'}\gg \fr{1}{R}$.
Only this kind of step by step compactification breaking of an initial
semisimple gauge group can give power law unification near the TeV
scale. It
would be also interesting to consider other extended gauge groups in
higher dimensions and to study the breaking pattern and the
phenomenological implications. We wish to study these and 
other relevant issues in our future publications.

\vspace{1cm}
{\bf Acknowledgement}

F.P.C. is supported by Funda\c c\~ ao de Ci\^ encia e Tecnologia (grant
SFRH/BD/4973/2001).

\section*{Appendix A: Influence of brane couplings \\
~~~~~~~ on gauge coupling unification}
\setcounter{equation}{0} 
\renewcommand{\theequation}{A.\arabic{equation}}

In this appendix we investigate the influence of some brane operators on 
gauge coupling renormalization. We will show that the 4D mass term of bulk
vectorlike states does not change the expressions for the power law
functions
$S$ and $S_1$, $S_2$ of (\ref{DeKKZ}) and (\ref{SKK}) resp.. Also,
it
turns out that some brane bi-linear derivative couplings, involving states
with negative parities, do not change this picture for $\mu_0\ll M\sim
M_G$. The latter condition is usually satisfied in models with 
power law unification
\footnote{Also in the models with additional
bulk vectorlike states, which we have studied, $\mu_0\ll
M_G$ is satisfied.}. 
Our discussion will be model independent. We will consider here the case
of 
$S^{(1)}/Z_2$ orbifold, but generalization to $S^{(1)}/Z_2\tm Z_2'$
scenarios is straightforward.

Consider two $N=2$ supermultiplets
${\bf E}_{N=2}=(E,~\ov{E})$ and ${\bf E'}_{N=2}=(E',~{\ov{E}}')$
transforming non trivially under a certain gauge group $G$.
If $E$, $E'$ belong to some representation ${\bf r}$ of $G$, then
$\ov{E}$, ${\ov{E}}'$ transform as ${\bf \bar r}$.
Let us assign the orbifold parities as

\beq
(E, ~{\ov{E}}')\sim {\bf +}~,~~~~(E',~\ov{E})\sim {\bf -}~.
\la{parA}
\eeq
Thus, only $E$, ${\ov{E}}'$ states have zero modes. Writing the brane
coupling
\begin{equation}
                     \int d^5xd^2\theta \de (y)\l 
\lambda_E\,E{\ov{E}}'+h.c.\r ~,
\la{brA}
\end{equation}
we obtain, after performing the integration over $y$, 
the 4D superpotential mass terms

\beq
W_{E}=2M_E\sum_{m, n}E^{(m)}{\ov{E}^{(n)}}'\eta^{(m)}\eta^{(n)}~,
\la{supA}
\eeq
where $M_E=\lam_E\mu_0/\pi =\lam_E/(\pi R)$ and $\eta^{(0)}=1/\sq{2}$,
$\eta^{(n)}|_{n\neq 0}=1$.
On the other hand, the last term in the 5D bulk 
Lagrangian (\ref{lagf}) gives 
direct mass terms for the KK
states

\beq
W^{KK}_E=\mu_0\sum_{n=1}^{\infty }n\l E^{(n)}\ov{E}^{(n)}+
{E^{(n)}}'{\ov{E}^{(n)}}'\r~.
\la{massKKA}
\eeq
{}For our calculations only a certain number of KK states is relevant,
which contribute
to the gauge coupling runnings. So, we truncate the tower of KK states on
the level $N_0$. Taking this into account and combining (\ref{supA}),
(\ref{massKKA}), the mass terms can be written in compact form as 
\begin{equation}
W_E+W^{KK}_E=\ov{{\mathcal E}}{\mathcal M}{\mathcal E} ~,
\la{matrixA}
\end{equation}
where
\begin{equation}
\ov{{\mathcal E}} =  
\left(~\bar{E}'^{(N)}~\ldots~\bar{E}'^{(0)}~\bar{E}^{(1)}~\ldots~
\bar{E}^{(N)}~\right),~~  {\mathcal E}^{T} = \left( ~E'^{(N)}~
\ldots~E'^{(1)}~E^{(0)}~\ldots~E^{(N)}~\right),
\end{equation}

\begin{equation}
                     {\mathcal M}\equiv\left[ \begin{array}{cc} 
\bar{{\mathcal M}}_0 & {\mathcal M}_E \\ 0 & {\mathcal M}_0 \end{array} 
\right]
,\quad {\mathcal M}_E = M_E \left[ \begin{array}{cccc} 
\sqrt{2} & 2 & \ldots & 2\\ \vdots & & \ddots & \vdots\\ \sqrt{2} & 2 & 
\ldots & 2 \\ 1 & \sqrt{2} & \ldots & \sqrt{2} \end{array} \right]~,
\la{matA}
\end{equation}
and
\begin{equation}                     
\bar{{\mathcal M}}_0= \mu_0 \left[ 
\begin{array}{cccc} 
N_0 & 0 & \ldots & 0 \\ 0 & N_0-1 &  & 0 \\ 
\vdots &  & \ddots & \vdots \\ 0 & 0 & \ldots & 1 \\ 
0 & 0 & \ldots & 0  \end{array} \right],
\qquad {\mathcal M}_0= \mu_0 
\left[ \begin{array}{ccccc} 
0 & 1 & \ldots & 0 & 0 \\ 
\vdots &  & \ddots & &\vdots \\ 
0 & 0 &  & N_0-1 & 0 \\ 
0 & 0 & \ldots & 0 & N_0 \end{array} \right]~.
\la{defmatA}
\end{equation}
To estimate the KK and zero mode contributions to the running, one should
diagonalize the matrix ${\mathcal M}$ and find the eigenvalues $M_0$,
$M_1$,
$\cdots $, $M_{2N_0}$. If all these masses lie below $M_G$, then their
contribution to the renormalization of $\al_i^{-1}$ up to the GUT scale
will be 
\begin{equation}
\De^{E}_i=-\fr{b_i^E}{2\pi }\sum_{n=0}^{2N_0}\ln \fr{M_G}{M_n}=
-\fr{b_i^E}{2\pi }\ln \fr{M_G^{2N_0+1}}{{\rm det }{\mathcal M}}~.
\la{DeA}
\end{equation}
{}From (\ref{matA}), (\ref{defmatA}) it is easy to see that
\beq
{\rm det }{\mathcal M}=M_E\l \prod_{n=1}^{N_0}M_n\r^2=
M_E\l \prod_{n=1}^{N_0}n\mu_0\r^2~.
\la{detMA}
\eeq
Using (\ref{detMA}) in (\ref{DeA}) we obtain
\beq
\De^E_i=-\fr{b^E_i}{2\pi }\ln \fr{M_G}{M_E}-
-\fr{\hat{b}^E_i}{2\pi }\sum_{n=1}^{N_0}\ln \fr{M_G}{n\mu_0}~,
\la{De1A}
\eeq
(where we have set $\hat{b}^E_i=2b^E_i$).
As one can see, the first term in (\ref{De1A}) is just the naive 
logarithmic running of
'zero mode' states, while the second term has precisely the form of
the 'unaffected' KK
power law running (\ref{DeKKZ}).
From this we conclude that due to the brane coupling (\ref{brA}),
the RGEs will in addition include a logarithmic term corresponding to
the zero
modes $E^{(0)}$, ${\ov{E}^{(0)}}'$ above the scale $M_E$, while the 
effect of the
KK states must be estimated by the power law function $S$. These
conclusions
are valid for any value of $M_E\stackrel{<}{_\sim }M_G$.

Higher order brane couplings, preserving orbifold symmetries, imply
$\pl_5$ derivatives and must be cut off at the fundamental scale $M\sim
M_G$.
For example, the couplings
\begin{equation}
\int d^5xd^2\theta \de (y)\left(
\frac{1}{M}E\partial_5\ov{E}+
\frac{1}{M}{\ov{E}}'\partial_5 E'\right)~, 
\la{dercoupA}
\end{equation}
after integration along the fifth dimension, change the masses by
the amount $\sim \mu_0^2/M$. The latter is negligible for $\mu_0\ll M$ and
(\ref{De1A}) is still valid.

\bibliographystyle{unsrt}

\newpage

\begin{figure}[h]
\vskip -3cm
\hglue -1cm
\epsfig{file=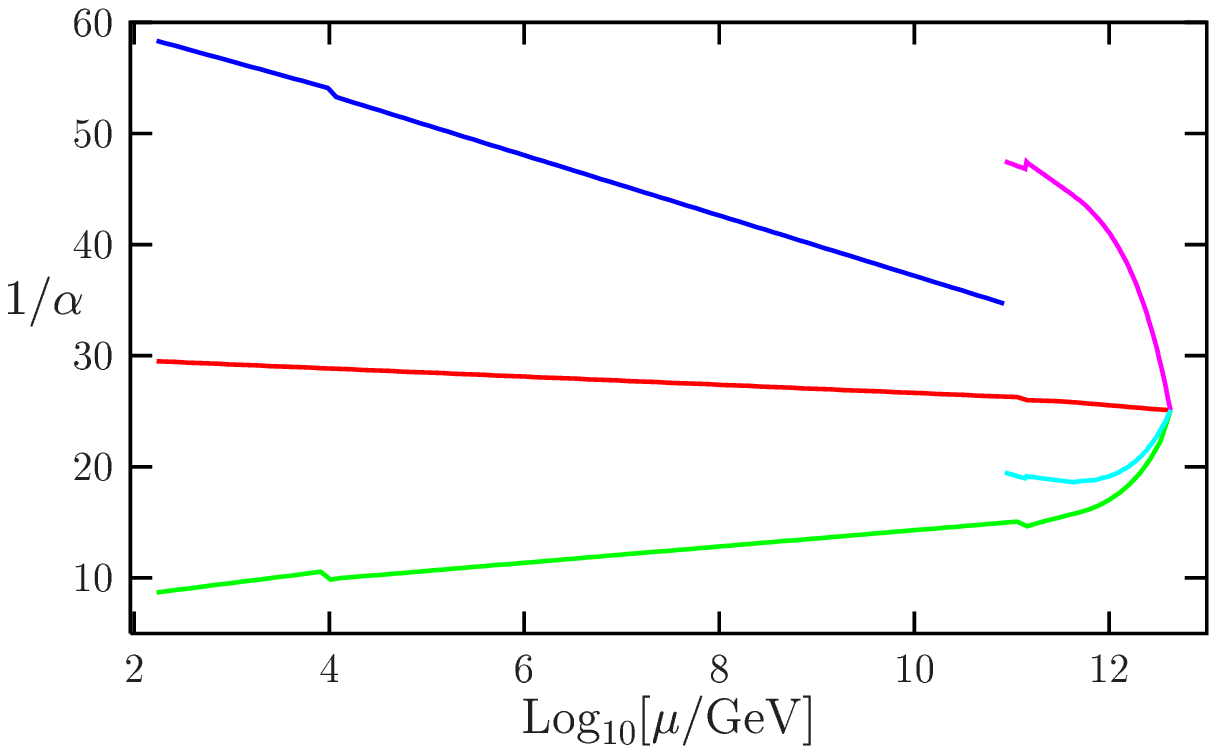,height=25cm,width= 16cm}
\vglue -14.5cm 
\hglue 10cm 
\vskip -0.5cm
\caption{\small 
Unification picture for model {\bf I}-susy422 
with $\al_s(M_Z)\simeq 0.119$ and $\eta=0$;
$M_I\simeq 8.5\cdot 10^{10}$~GeV, 
$\mu_0\simeq 1.4\cdot 10^{11}$~GeV,
$M_G\simeq 4.3\cdot 10^{12}$~GeV.
}
\label{sur1}
\end{figure}

\vglue -5cm 
\vspace{-1cm}

\begin{figure}[b]
\vskip -15cm
\hglue -1cm
\epsfig{file=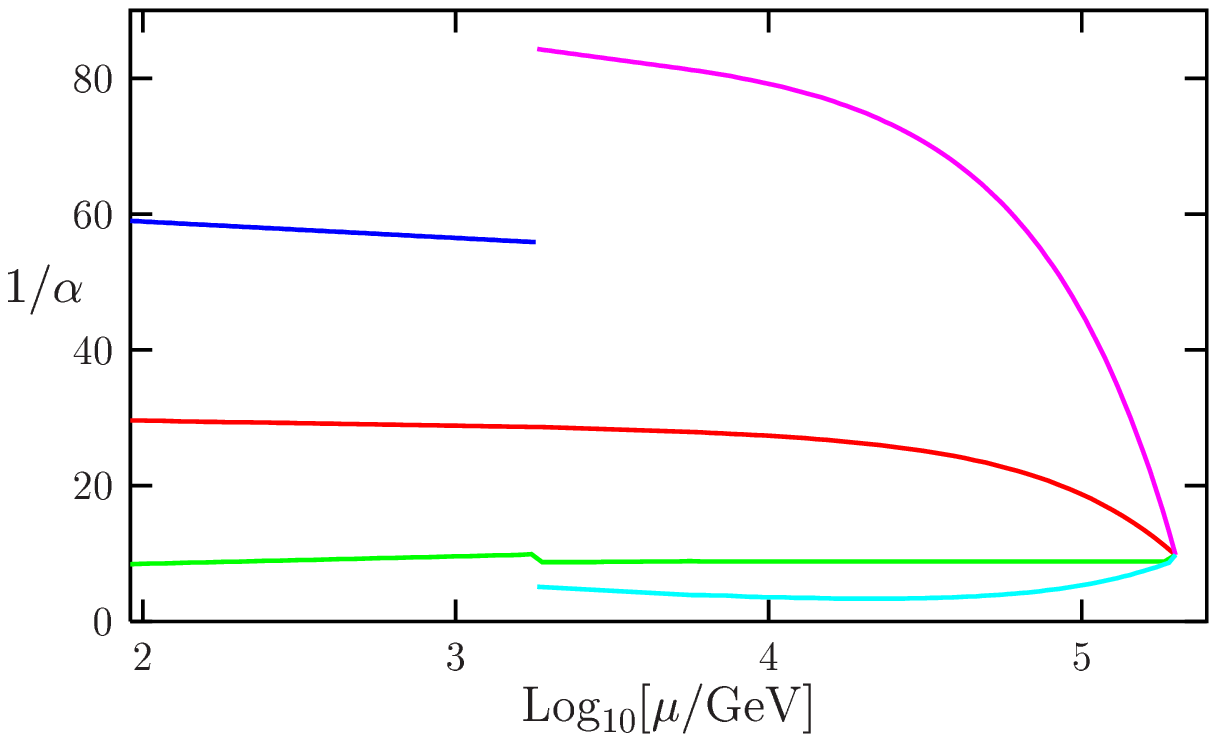,height=25cm,width= 16cm}
\vglue -14.5cm 
\hglue 10cm 
\vskip -0.3cm
\caption{\small 
Low scale unification picture for model {\bf III}-susy422 with
$\al_s(M_Z)\simeq 0.1184$ and $\eta=0$;
$M_I=\mu_0\simeq 1.8$~TeV,
$M_G\simeq 200$~TeV.
}
\label{sur2}
\end{figure}

\newpage

\begin{figure}[h]
\vskip -3cm
\hglue -1cm
\epsfig{file=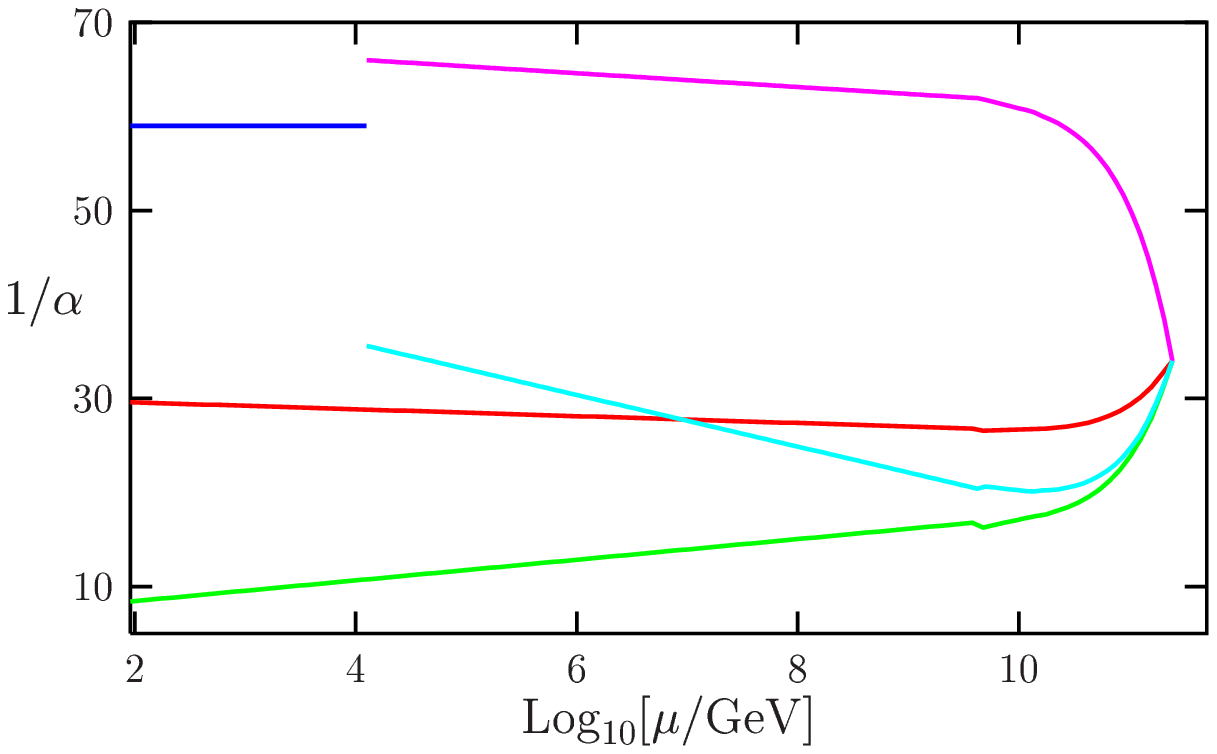,height=25cm,width= 16cm}
\vglue -14.5cm 
\hglue 10cm 
\vskip -0.5cm
\caption{\small 
Unification picture for Model {\bf I'}-susy422 with $\al_s(M_Z)\simeq
0.119$ and $\eta =0$;
$M_I\simeq 12.6$~TeV, 
$\mu_0\simeq 4.4\cdot 10^{9}$~GeV,
$M_G\simeq 2.5\cdot 10^{11}$~GeV.
}
\label{sur3}
\end{figure}

\vglue -5cm 
\vspace{-1cm}

\begin{figure}[b]
\vskip -15cm
\hglue -1cm
\epsfig{file=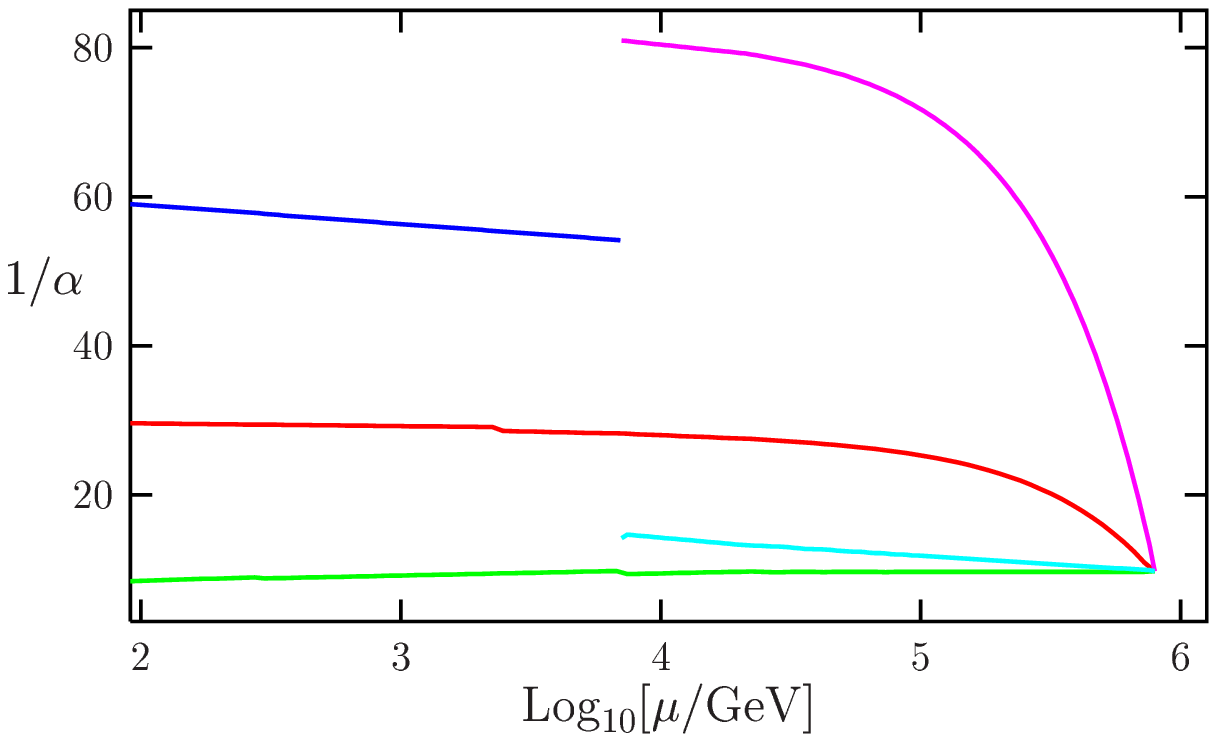,height=25cm,width= 16cm}
\vglue -14.5cm 
\hglue 10cm 
\vskip -0.3cm
\caption{\small 
Low scale unification picture for Model {\bf II'}-susy422
with $\al_s(M_Z)\simeq 0.119$ and $\eta=0$;
$M_I\simeq 7.1$~TeV, 
$\mu_0\simeq 7.2$~TeV,
$M_G\simeq 794$~TeV.
}
\label{sur4}
\end{figure}

\newpage

\begin{figure}[h]
\vskip -3cm
\hglue -1cm
\epsfig{file=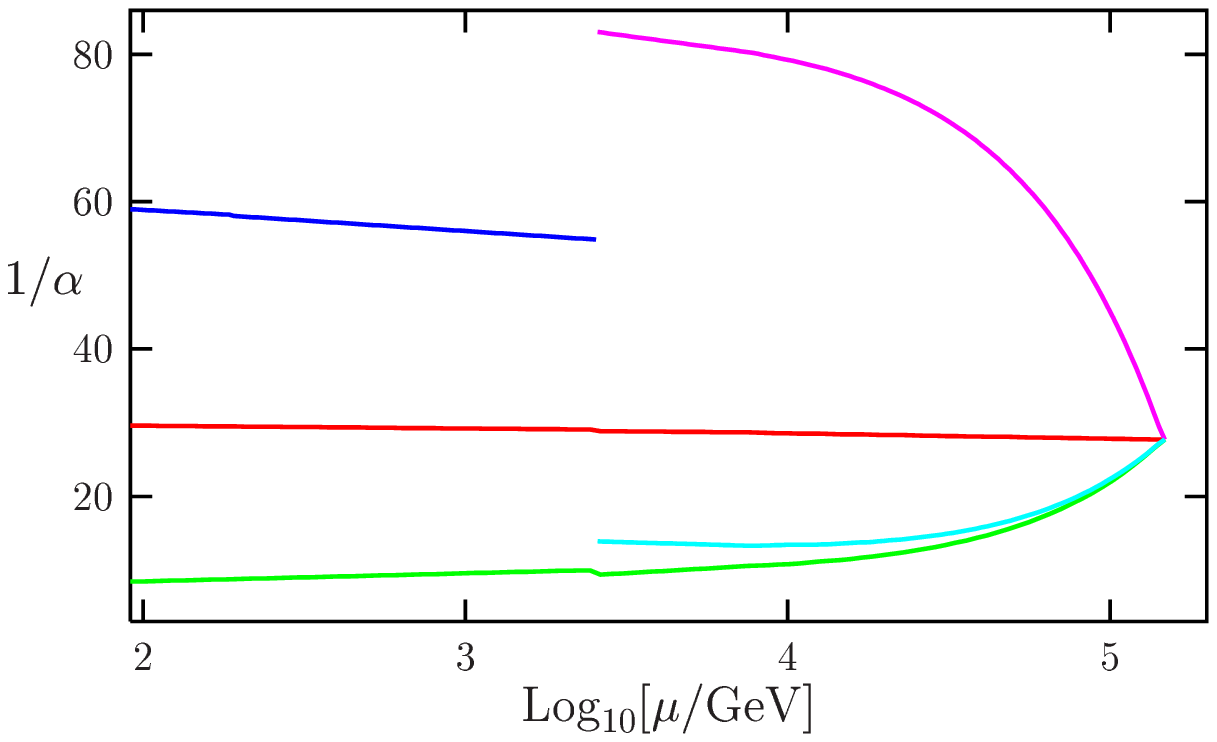,height=25cm,width= 16cm}
\vglue -14.5cm 
\hglue 10cm 
\vskip -0.5cm
\caption{\small 
Low scale unification picture for model {\bf III'}-susy422 
with $\al_s(M_Z)\simeq 0.119$  and  $\eta=0$;
$M_I=\mu_0\simeq 2.6$~TeV,
$M_G\simeq 148$~TeV.
}
\label{sur5}
\end{figure}

\vglue -5cm 
\vspace{-1cm}

\begin{figure}[b]
\vskip -15cm
\hglue -1cm
\epsfig{file=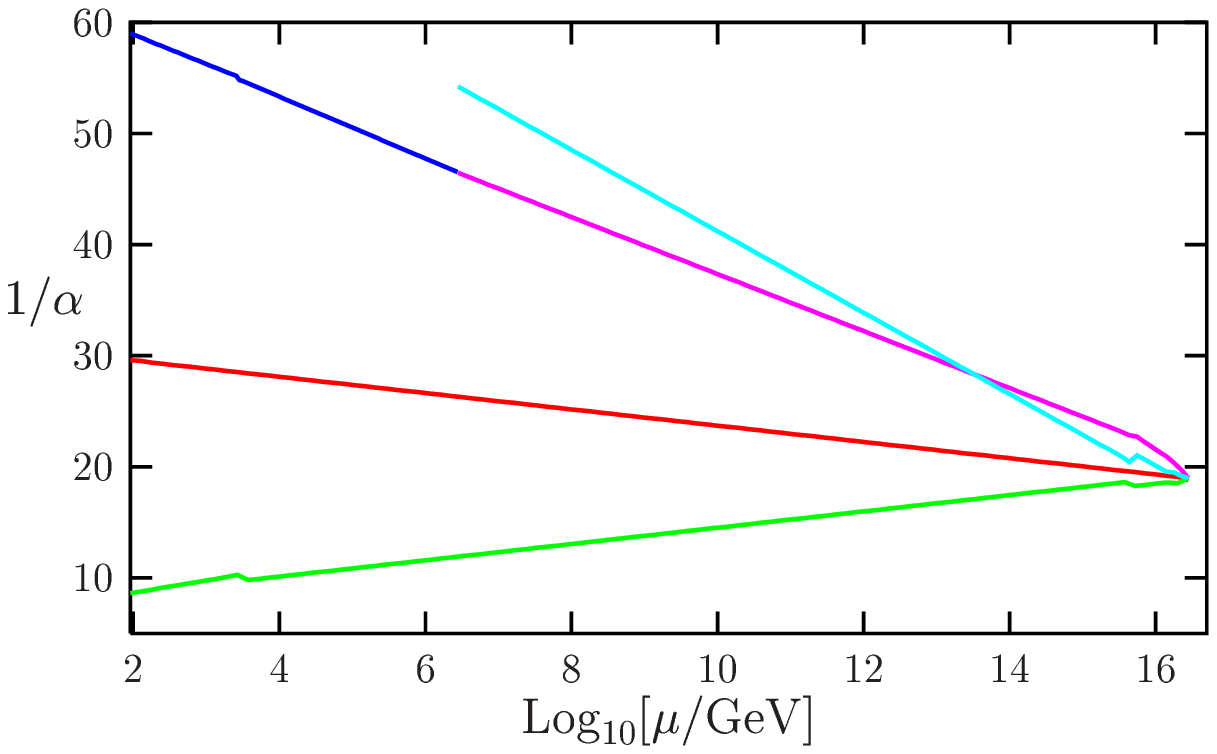,height=25cm,width= 16cm}
\vglue -14.5cm 
\hglue 10cm 
\vskip -0.3cm
\caption{\small 
Unification picture for flipped $SU(5)\tm U(1)$ model
with $\al_s(M_Z)\simeq 0.116$ and  $\eta=0$;
$M_I\simeq 2.8\cdot 10^{6}$~GeV, 
$\mu_0\simeq 4.7\cdot 10^{15}$~GeV,
$M_G\simeq 2.8\cdot 10^{16}$~GeV. 
}
\label{sur6}
\end{figure}

\end{document}